\pdfoutput=1

\documentclass{ws-ijmpd-dac}
\usepackage[super,compress]{cite}

\usepackage{amsfonts,amssymb,amsmath}

\usepackage{ifthen}
\makeatletter%
\usepackage{xstring}
\IfSubStr{\@classoptionslist}{preprint}%
{}%
{}%
\makeatother%

\newcommand{\arxiv}{1}

\usepackage{graphicx} 
\usepackage[caption=false]{subfig}             

\graphicspath{{Figures/}}

\usepackage[breaklinks]{hyperref}
\hypersetup{colorlinks,urlcolor=black,citecolor=black,linkcolor=black,filecolor=black}
\usepackage{breakurl}

\usepackage{enumerate} 
\usepackage{enumitem} 

\newcommand{\ket}[1]{\mbox{$| {#1} \rangle$}}
\newcommand{\bracket}[2]{\mbox{$\langle {#1} \!\mid\! {#2} \rangle$}}
\newcommand{\ketbra}[2]{\mbox{$| {#1} \rangle\langle {#2} |$}}
\newcommand{\melt}[3]{\mbox{$\langle {#1} | {#2} | {#3} \rangle$}}

\newcommand{\expct}[1]{\mbox{$\langle {#1} \rangle$}}

\newcommand{\Projsupb}[2]{\mbox{$P^{{#1}}_{\mbox{\scriptsize{${#2}$}}}$}}

\newcommand{\tr}[1]{\mbox{${\rm tr}[ {#1} ]$}}

\newcommand{\tri}[2]{\mbox{${\rm tr}[ {#1} \rho {#2} ]$}}

\def\Re{\mathbb{R}}

\def\Int{\mathbb{Z}}

\usepackage{dsfont}
\def\Id{\mathds{1}}

\def\sys{{\mathcal S}}

\def\cell{{\mathcal V}}

\def\ow{\mathring{\omega}}
\def\oV{\mathring{V}}
\def\oe{\mathring{e}}
\def\oq{\mathring{q}}

\def\lp{{l}_{p}}

\def\cons{\sqrt{12 \pi G}}
\def\k{\kappa}

\def\v{\nu}

\def\rhom{\rho_{\mathrm{max}}}

\def\sWdW{{\scriptscriptstyle\text{WdW}}}
\def\sLQC{{\scriptscriptstyle\text{LQC}}}

\def\eW{e^{\sWdW}}
\def\UW{U^{\sWdW}} 
\def\eS{e^{{\scriptscriptstyle (s)}}}

\def\Psik{\tilde{\Psi}}

\def\PsiW{\Psi^{\sWdW}}
\def\PhiW{\Phi^{\sWdW}}
\def\PsiWL{\Psi^{\sWdW}_{L}}
\def\PsiWR{\Psi^{\sWdW}_{R}}
\def\PsiWk{\tilde{\Psi}^{\sWdW}}

\def\TW{\Theta^{{\scriptscriptstyle\text{WdW}}}}

\begin{document}


\title{THE CONSISTENT HISTORIES APPROACH TO LOOP QUANTUM COSMOLOGY}

\author{David A.~Craig}

\address{%
Department of Chemistry and Physics, Le Moyne College\\
Syracuse, New York, 13214, USA\\
craigda@lemoyne.edu
}

\maketitle

\begin{history}
\received{Day Month Year}
\revised{Day Month Year}
\end{history}

\begin{abstract}
We review the application of the consistent (or decoherent) histories
formulation of quantum theory to canonical loop quantum cosmology.
Conventional quantum theory relies crucially on ``measurements'' to convert
unrealized quantum potentialities into physical outcomes that can be assigned
probabilities.  In the early universe and other physical contexts in which
there are no observers or measuring apparatus (or indeed, in any closed
quantum system), what criteria determine which alternative outcomes may be
realized and what their probabilities are?  In the consistent histories
formulation it is the vanishing of interference between the branch wave
functions describing alternative histories -- as determined by the system's
\emph{decoherence functional} -- that determines which alternatives may be
assigned probabilities.  We describe the consistent histories formulation and
how it may be applied to canonical loop quantum cosmology, describing in
detail the application to homogeneous and isotropic cosmological models with
scalar matter.  We show how the theory may be used to make definite physical
predictions in the absence of ``observers''.  As an application, we
demonstrate how the theory predicts that loop quantum models ``bounce'' from
large volume to large volume, while conventional ``Wheeler-DeWitt''-quantized
universes are invariably singular.  We also briefly indicate the relation to
other work.
\end{abstract}


\ccode{PACS numbers: 98.80.Qc,04.60.Pp,03.65.Yz,04.60.Ds,04.60.Kz}  


\section{Introduction}
\label{sec:intro}

The quantum mechanics employed by physicists every day to make predictions
concerning the outcomes of experimental measurements is not, as it is
conventionally understood, sufficient to make predictions in physical
situations in which a clearly defined concept of ``measurement'' is absent.
Prime examples of such physical situations may be found in the early universe,
or indeed, any time or place in the universe where measuring (recording)
apparatus are not present.  How, then, may physical theories offer clear and
unambiguous predictions concerning the behavior of the physical universe in
the moments surrounding the epoch normally considered to be the big bang?

One (possibly partial) answer to this question lies in the consistent (or
decoherent) histories formulation of quantum theory pioneered and subsequently
developed by
Griffiths \cite{griffiths08}, 
Omnes \cite{omnes94}, 
Gell-Mann and Hartle \cite{GMH90a,GMH90b,hartle91a,lesH}, 
Halliwell \cite{halliwell99,hallithor01,hallithor02,halliwall06,halliwell09} 
and others.\cite{hartlemarolf97,CH04,as05}  In this framework quantum theory is
supplemented by a \emph{consistency condition} which must be satisfied in any
specific physical situation in order for quantum theory to offer definite
predictions concerning that situation.  This condition amounts to the
requirement that there is essentially no interference (overlap) between the
branch wave functions corresponding to each of the 
alternative quantum histories describing that scenario.  Only in that case can
quantum theory consistently assign definite probabilities to each of those
possible
alternative histories.  This condition of ``decoherence''%
\footnote{This usage of the term ``decoherence'' is related to, but
conceptually distinct from, the broadly understood phenomenon of
\emph{environmental
decoherence}.\cite{giulini,schlosshauer07,halliwell89,zurek09a,RZZ16a} Physical
mechanisms which engender environmental decoherence may thereby lead to the
decoherence of the corresponding histories.  However, other physical processes
such as occur in typical laboratory ``measurement situations'' may also lead
to decoherence of histories in the sense the term is applied here.
} %
or ``consistency'' among the branch wave functions is satisfied in any
physical scenario that would normally be thought of as a classical
``measurement situation'',%
\footnote{This is epitomized by the two-slit experiment.  When the slit through
which the particle passes is measured, in the sense that sufficient
information is recorded to determine it, the interference between the branch
wave functions corresponding to passage through either the upper or lower slit
separately is destroyed, and physically meaningful probabilities can be 
assigned to which slit the particle traversed.  Otherwise, the alternative 
wave functions interfere, and quantum mechanics simply has nothing to say -- 
i.e.\ cannot assign a probability in a logically consistent manner -- about 
through which slit the particle passed.
} %
and hence the framework of consistent histories encompasses the ordinary
quantum mechanics of measurements -- sans the additional hypothesis of
``collapse of the wave function'' upon measurement.  However, it also extends
it in an objective, observer-independent manner to a wide range of physical
situations in which observers and measuring apparatus are not present, thus
giving quantum theory a voice even when and where physicists themselves do not
have one.

The central mathematical object in the consistent histories framework is the
\emph{decoherence functional}.  It is a natural generalization of the notion
of the quantum state of a system as it arises in the algebraic formulation of
quantum theory, and is a sesquilinear functional of the possible quantum
histories of the system.  The decoherence functional measures the quantum
interference between the possible histories, and, when that interference
vanishes among all the possible histories -- i.e.\ the family of histories
\emph{decoheres} or ``is consistent'' -- the probabilities of each of them.
In this way the decoherence functional -- the quantum state itself -- provides
definite quantum predictions for the behavior of a system even in the absence
of observers, recording apparatus, or ``measurements''.%
\footnote{What the consistent histories framework does \emph{not} dictate is
which family of quantum histories one interrogates.  Different choices of
family may provide very different pictures of ``what happens''.  This, indeed,
is a manifestation of contextuality in the quantum mechanics of history --
quantum mechanics is irreducibly a contextual theory, and thereby paints a
contextual picture of reality.
} %

In this contribution we will summarize recent work on the application of these
ideas to loop quantum cosmology, developed in collaboration with Parampreet
Singh,\cite{CS10a,CS10b,CS10c,CS12a,CS13a,CS16c} and based on earlier work of
J.B.\ Hartle and
others.\cite{hartle91a,lesH,halliwell99,hallithor01,hallithor02,halliwall06,halliwell09,hartlemarolf97,CH04,as05}
All of the discussion will be formulated in the context of a flat, homogeneous
and isotropic Friedmann-Lema\^{i}tre-Robertson-Walker (FLRW) cosmological
model sourced by a massless, minimally coupled scalar field.  In section
\ref{sec:gqt} we describe the consistent histories formalism in general terms.
We then discuss in section \ref{sec:chqc} the implementation of this formalism
in both a conventional so-called ``Wheeler-DeWitt'' quantization of this
cosmology, as well as a distinct \emph{loop} quantization of the same model.
In section \ref{sec:app} we illustrate the application of this formalism to
show how it may be used to make predictions concerning quantum histories of
various physical quantities.  As an important illustration, we show that in
the Wheeler-DeWitt quantization, all quantum states are invariably singular,
as they are in the classical theory.  By contrast, in the loop quantization
all quantum states remain non-singular, and indeed ``bounce'' from large
volume to large volume, approaching well-defined states of the corresponding
Wheeler-DeWitt quantization.  In this way we illustrate how consistent
histories formulations of quantum cosmological theories may be defined and
then deployed to make consistent quantum predictions in these theories in the
absence of the observers or measurements typically viewed as essential to
prediction in quantum mechanics.

\section{Generalized Consistent Histories Quantum Theory}
\label{sec:gqt}

A ``generalized quantum mechanics'', as originally defined by
Hartle\cite{hartle91a,lesH}, consists in the specification of four
ingredients: (i) The \emph{\textbf{fine-grained histories}} $h$ of the system,
the most refined descriptions of the possible alternative physical histories
of the system it is possible to give.  In Lagrangian mechanics, for example,
for a set of generalized coordinates $\{q_i\}$, the fine-grained histories are
all possible paths $\{q_i(t)\}$.  The fine-grained histories may be collected
into exhaustive (complete) sets of mutually exclusive alternative histories
which we will sometimes denote by $\sys =\{h\}$.  Depending upon the theory,
there may be many distinct (possibly incompatible) such exclusive, exhaustive
sets of histories, as happens for example in ordinary Hilbert space quantum
mechanics because of the existence of non-commuting operators.  (Explicit
examples of histories in Hilbert space quantum theory will be given below.)
(ii) The allowed \emph{\textbf{coarse-grainings}} (partitions) of exclusive,
exhaustive sets of alternative histories into more coarse-grained descriptions
of the system.  For example, in a diffeomorphism-invariant theory it would be
typical to require coarse-grainings to be themselves diffeomorphism-invariant.
(It is common to denote by $\bar{h}$ a course-grained history that contains
the history $h$ when it is convenient to refer back to $h$ specifically.  It
is also common to allow $h$ alone to represent both fine-grained and
coarse-grained histories as is convenient in context.)  It is assumed that all
exclusive, exhaustive sets of fine-grained histories $\sys =\{h\}$ have a
common complete coarse-graining $u=\cup_{h\in\sys}\,h$.  (iii) The
\emph{\textbf{decoherence functional}} $d$.  The decoherence functional is a
complex-valued function on pairs of histories that is
\begin{description}
\item[(i)] \emph{Hermitian}: \quad $d(h,h')=d(h',h)^*$

\item[(ii)] \emph{Positive}: \quad  $d(h,h)\geq 0$

\item[(iii)] \emph{Additive} (``Principle of Superposition''): \quad
         $d(\bar{h},\bar{h}') = \sum_{h\in\bar{h}}\sum_{h'\in\bar{h'}} d(h,h')$

\item[(iv)] \emph{Normalized}:  \quad $\sum_{h,h'\in\sys}d(h,h')=1$
\end{description}
Note these conditions imply $d(u,u)=1$.   
Finally, the decoherence functional is used to define (iv) the 
\emph{\textbf{decoherence condition}} that determines when probabilities may 
be assigned to the histories in an exclusive, exhaustive set $\sys$.  
The simplest and most common decoherence condition (usually called ``medium 
decoherence'' in the literature\cite{GMH93,diosi04}) simply requires that the 
decoherence functional be diagonal on a set of histories $\sys$ in 
order for probabilities to be assigned to the histories in $\sys$.  In 
that case the diagonal elements of $d$ are the corresponding probabilities, 
and $\sys$ is said to be a \emph{consistent} or \emph{decoherent} set 
of histories:
\begin{equation}
d(h,h') = p(h)\, \delta_{h,h'}   \qquad \forall h,h'\in\sys
\label{eq:dfndmtl}
\end{equation}
if $\sys$ is a consistent set.  Only when $\sys$ is consistent
may the diagonal elements of $d$ be interpreted as (Kolmogorov) probabilities,
for which $p(h)\geq 0$ and $\sum_{h\in\sys} p(h)=1$ (whence the term
``consistent histories''.)  In particular, if decoherence does not obtain, the
putative ``probabilities'' $p(h)$ do not add consistently in the sense that
$p(h_1+h_2)\neq p(h_1) + p(h_2)$, as happens for example in the two-slit
experiment when the slit traversed is not measured.

It is the consistency condition (\ref{eq:dfndmtl}) that alone determines
whether or not probabilities may be meaningfully assigned in any family
of histories, and not any notion of ``observer'' or ``measurement'' (though it
does reproduce the predictions of ordinary Copenhagan quantum theory when
quantum measuring apparatus are included in the system.)  It is an objective
criterion that depends only on the system's quantum state and dynamics -- and,
of course, on the set of histories in question.

In ordinary Hilbert space quantum theory, fine-grained histories may be 
specified by sequences  of eigenvalues $a^{\alpha_i}_{k_i}$ of operators 
$A^{\alpha_i}$ at a sequence of times $\{t_i\}$: 
$h=(a^{\alpha_1}_{k_1},a^{\alpha_2}_{k_2},\cdots)$.%
\footnote{To help clarify the confusing but necessary notation, note that 
the spectrum of the operator $A^{\alpha}$ is $\{\cup_k a^{\alpha}_k\}$.  In 
other words, the \emph{superscript} labels the \emph{observable}, 
while the \emph{subscript} labels the \emph{eigenvalues} of that observable.
} %
Corresponding to each history $h$ may be defined the operator 
\begin{equation}
C_h = \Projsupb{\alpha_1}{a_{k_1}}(t_1)
        \Projsupb{\alpha_2}{a_{k_2}}(t_2) \cdots
        \Projsupb{\alpha_n}{a_{k_n}}(t_n),
\label{eq:classopdef-fg}
\end{equation}
where if $U(t)$ is the theory's propagator,
$P_a(t_i)=U(t_i,t_0)^{\dagger}\ketbra{a}{a}U(t_i,t_0)$ is a Heisenberg picture
projection operator.  $C_h$ is called the ``class operator'' for
the history $h$, and is typically identified with $h$ itself.
Coarse-grainings of these histories then correspond to the operator sums
\begin{equation}
C_{\bar{h}} = 
\sum_{a_{k_{1}}\in\Delta a_{\bar{k}_1}}
\sum_{a_{k_{2}}\in\Delta a_{\bar{k}_2}}\cdots 
\sum_{a_{k_{n}}\in\Delta a_{\bar{k}_n}} \, 
  \Projsupb{\alpha_1}{a_{k_{1}}}(t_1)
  \Projsupb{\alpha_2}{a_{k_{2}}}(t_2) \cdots
  \Projsupb{\alpha_n}{a_{k_{n}}}(t_n)
\label{eq:classopdef-cg}
\end{equation}
corresponding to the history $\bar{h}=(\Delta a^{\alpha_1}_{\bar{k}_1},\Delta
a^{\alpha_2}_{\bar{k}_2},\cdots)$ in which the value of $A^{\alpha_1}$ is in the
range $\Delta a^{\alpha_1}_{\bar{k}_1}$ labeled by $\bar{k}_1$ (so the spectrum of
$A^{\alpha}$ is the union of the intervals $\{\cup_k \Delta a^{\alpha}_{k}
\}$), and so on.  The completely coarse-grained history $u$
then corresponds to
\begin{equation}
C_u = \sum_{h\in\sys}C_h = \Id.
\label{eq:classopdefCu}
\end{equation}
The decoherence functional corresponding to ordinary quantum mechanics is
\begin{equation}
d(h,h') = \tri{C_h^{\dagger}}{C_{h'}},
\label{eq:dfdefqm}
\end{equation}
where $\rho$ is the initial density matrix.%
\footnote{In this form it should be clear in what sense the decoherence
functional is a natural generalization of the notion of the ``quantum state''
of the system as it arises in the algebraic formulation of quantum theory, to
measure interference between histories as well as their
probabilities.\cite{ILS94a,dac97}
} %
If the intial state is pure, $\rho=\ketbra{\psi}{\psi}$, this reduces simply 
to
\begin{equation}
d(h,h') = \bracket{\psi_{h'}}{\psi_h},
\label{eq:dfdefstd}
\end{equation}
where
\begin{equation}
\ket{\psi_h(t)} = U(t,t_0)C_h^{\dagger}\ket{\psi}
\label{eq:bwfdef}
\end{equation}
is the so-called ``branch wave function'' corresponding to the history $h$.
Up to normalization, it is simply the wave function at time $t$ a system that
began in the state $\ket{\psi}$ would have were the system observed to have
had values of observable $A^{\alpha_1}$ in $\Delta a^{\alpha_1}_{k_1}$ at time
$t_1$, %
$A^{\alpha_2}$ in $\Delta a^{\alpha_2}_{k_2}$ at time $t_2$, %
and so on, i.e.\ for the system to have been observed to ``follow'' the
particular history $h$.  However, in consistent histories quantum theory we do
\emph{not} (necessarily) have observers as an essential element of the
predictive scheme, and consequent wave function ``collapse''.  Here, it
appears simply as a description of one of the many possible alternative
histories of the system, and it is up to decoherence to decide if the
corresponding family of histories is consistent, and if so, what the 
probability of each such decohering history may be.%
\footnote{See Refs.~\refcite{CS10c,CS13a} for some discussion of the leading 
factor of the propagator $U$, which is not strictly necessary -- it cancels 
in the decoherence functional -- and is introduced only in order to enable us 
to think of $\ket{\psi_h}$ as an evolving solution of the Schr\"{o}dinger 
equation for all $t$. 
} %

If there is also a ``final condition'' $\rho_{\omega}$ in addition to an 
``initial condition'' $\rho_{\alpha}$, then the decoherence functional may be 
defined similarly by 
$d(h,h') = \tr{\rho_{\omega}C_h^{\dagger}\rho_{\alpha}C_{h'}}$.    If 
$\rho_{\alpha}=\sum_i \alpha_i\ketbra{\Psi_i}{\Psi_i}$ and
$\rho_{\omega}=\sum_i \omega_i\ketbra{\Phi_i}{\Phi_i}$, this reduces to
\begin{equation}
d(h,h') = \sum_{i,j}\alpha_i\omega_j \melt{\Psi_i}{C_h}{\Phi_j}^*
              \melt{\Psi_i}{C_{h'}}{\Phi_j}.
\label{eq:dfdefTS}
\end{equation}

Finally, given a \emph{path integral} formulation of a quantum theory, if the 
propagator is given by
\begin{equation}
U(t,t_0) = \int\delta q\, e^{i\int_{t_0}^{t}dt\, S[q]},
\label{eq:propdefPI}
\end{equation}
for some action $S[q]$, then class operators may be defined by
\begin{equation}
C_h^{\dagger}(t,t_0) = \int_{q(t)\in h}\delta q\, e^{i\int_{t_0}^{t}dt\, 
S[q]}.
\label{eq:classopdefPI}
\end{equation}
In other words, in a path integral formulation, coarse-grained histories are
defined by partitions of the space of paths which appear in the path integral.
Considerable further detail concerning the construction of the decoherence
functional in theories whose amplitudes are defined by path integrals, both
relativistic and non-relativistic, may be found in
Refs.~\refcite{hartle91a,lesH,hallithor01,hallithor02,halliwall06,halliwell09,CH04,CS13b}.
In this way, in conjunction with Eq.\ (\ref{eq:dfdefTS}), a consistent
histories formulation may be given to spin foam models of quantum
gravity\cite{schroer13a} and loop quantum cosmology,\cite{CS13b} though we
will not discuss these here.

\subsection{Prediction in consistent histories quantum theory}
\label{sec:prediction}

The process of making a physical prediction in this formulation of quantum 
theory proceeds as follows.  Given a quantum state $\ket{\psi}$ and its 
associated decoherence functional, one constructs the class operators for the 
family of histories corresponding to the physical question in which one is 
interested -- for example, what (coarse-grained) path does a quantum particle 
follow through a two-slit apparatus.  The decoherence functional is then 
calculated to determine whether or not that family of histories decoheres 
i.e.\ is consistent.  If the family is consistent, the decoherence functional 
determines the probability of each history in the family, and the quantum 
question has been answered.

However, if the family of histories is \emph{not} consistent, then no
probabilities can be assigned, and quantum mechanics says that the question
being asked \emph{has no consistent answer within the theory.} Familiar
examples are replete in ordinary quantum theory -- e.g.\ the precise values of
the position and momentum of a particle at the same moment, or which slit a
particle passed through in the two slit experiment if no information was
gathered to determine it.  This important point is perhaps worth some
emphasis: the inability of quantum mechanics to make definite predictions
concerning what are classically perfectly sensible physical statements is not
in any way a unique feature of consistent histories quantum theory.  What the
consistent histories formulation provides is simply an
\emph{observer-independent criterion}, derived from the quantum state itself,
for determining when that is the case, rather than relying on the Copenhagen
notion of a wave function-collapsing ``measurement'', which is in any case
impossible to make sense of in environments such as the early universe.

We will see how this scheme works in quantum cosmology in the
sequel.

\section{Homogeneous and Isotropic Scalar Cosmologies}
\label{sec:hicosmo}

We consider what must be the simplest possible cosmological model, namely, a
flat, homogeneous and isotropic Friedmann-Lema\^{i}tre-Robertson-Walker
cosmology sourced by a single massless, minimally coupled scalar field.  This
model has two relevant advantages.  First, it is simple enough that it admits
two distinct, exactly solvable quantizations: a conventional
``Wheeler-DeWitt'' quantization, and a \emph{unitarily inequivalent} loop
quantization, the exactly solvable version of which is dubbed sLQC. Moreover,
the loop quantization can be cast in the form of a spin-foam (quasi-``path
integral'') quantization of flat scalar FLRW, with an explicitly solvable
vertex expansion.  It is thus a perfect playground within which to explore and
contrast the consistent histories formulations of each of these three
quantizations in a context in which mathematically explicit expressions and
exact calculations are possible.  Second, in this model the scalar field
emerges as an internal physical ``clock''.  While not essential to the
consistent histories formulation, the presence of an internal clock variable
does make it easier to conceptualize and interpret the results.
Thus, this model is rich enough to supply an exceptional proving ground for 
the exploration of the technical and conceptual issues associated with the 
consistent histories formulation of quantum gravitational models, in a 
technically manageable context.

\subsection{Classical scalar cosmologies}
\label{sec:classcosmo}

The FLRW metrric for a homogeneous isotropic universe may be expressed as
\begin{equation}
g_{ab} = - N(t)^2dt_a dt_b + a(t)^2 \oq_{ab},
\label{eq:FLRWmetric}
\end{equation}
where $a(t)$ is the scale factor, $N(t)$ is the lapse, and $\oq_{ab}$ is a 
fixed, flat fiducial metric on the spatial slices $\Sigma$, which we take to 
be topologically $\Re^3$.\cite{ashsingh11}  To construct a Hamiltonian 
formulation of the dynamics spatial integrals over finite volumes are 
required.  Therefore, one introduces a fixed fiducial cell $\cell$ with 
volume $\oV$ relative to $\oq_{ab}$, the physical volume of which is 
therefore $V=a^3\oV$.  The Einstein-Hilbert action for a flat ($k=0$) FLRW 
universe sourced by a massless, minimally coupled scalar field becomes, after 
integration over the fiducial spatial volume $\cell$,
\begin{equation}
S = \oV\int dt\, \left\{ 
-\frac{3}{8\pi G}\frac{a\dot{a}^2}{N} + \frac{1}{2}a^3\frac{\dot{\phi}^2}{N}
\right\} ,
\label{eq:FLRWaction}
\end{equation}
where the dot denotes a derivative with respect to $t$.  Choosing phase space 
variables with Poisson brackets $\{a,p_a\}=1$ and $\{\phi,p_{\phi}\}=1$, the 
canonical momenta are
\begin{equation}
p_a = -\frac{3}{4\pi G}\oV a \frac{\dot{a}}{N},   \qquad 
p_{\phi} = \oV a^3 \frac{\dot{\phi}}{N}.
\label{eq:FLRWmom}
\end{equation}
The Hamiltonian is then
\begin{equation}
H = \frac{1}{\oV}\left\{
-\frac{2\pi G}{3}\frac{N}{a}p_a^2 + \frac{1}{2}\frac{N}{a^3}p_{\phi}^2
\right\}.
\label{eq:FLRWHam}
\end{equation}
For comparison with loop quantum cosmology it turns out to be convenient to 
make a canonical transformation to a different set of variables, a volume 
variable%
\footnote{Note we have adopted the conventions used in
Refs.~\refcite{dac13a,CS13a}, which differ slightly from the normalization
chosen in Ref.\ \refcite{CS10c}.
} %
\begin{equation}
\v = \varepsilon\, \frac{1}{2\pi\lp^2}\,\frac{\oV}{\gamma}\,a^3
\label{eq:nudef}
\end{equation}
and 
\begin{eqnarray}
b & = & -\varepsilon\, \frac{4\pi G}{3}\, \frac{\gamma}{\oV}\,\frac{p_a}{a^2}
\nonumber\\
 & = & \varepsilon\, \gamma\, \frac{\dot{a}/N}{a}.
\label{eq:bdef}
\end{eqnarray}
Here $G$ is Newton's constant, $\lp = \sqrt{G\hbar}$ is the Planck length 
(with $c=1$), and $\gamma$ is the Barbero-Immirzi parameter of loop quantum 
gravity; its presence is purely for convenience of comparison to LQC and it 
can otherwise be set to 1.  $\varepsilon = \pm 1$ is a factor that determines 
the orientation relative to a fiducial triad in terms of which $\oq_{ab}$ is 
expressed.  Its presence is necessary for a consistent quantization because 
it ensures $-\infty < \v < \infty$ as  well as  $-\infty < b < \infty$.  
These variables satisfy  $\{b,\v\}=2/\hbar$.
Physically, the volume of $\cell$ is $V=2\pi\lp^2\gamma|\v|$, and from
(\ref{eq:bdef}), $b$ is $\varepsilon\gamma\times (\mathrm{Hubble\ rate})$.
Equivalently, since for flat FLRW the Ricci scalar is
$R=-6(\dot{a}/Na)^2=-6(b/\gamma)^2$, $b$ is also a measure of spacetime
curvature.

The classical dynamical trajectories given by the solution of Hamilton's 
equations show that $p_{\phi}$ is a constant of the motion, and
\begin{equation}
\phi = \pm \k^{-1}\, \ln\left|\frac{\v}{\v_0}\right| + \phi_0, \qquad 
V = V_0 e^{\pm\k (\phi-\phi_0)},
\label{eq:FLRWclasssoln}
\end{equation}
where $\k=\cons$ and $\v_0$ and $\phi_0$ are constants of integration.  Thus
the classical solutions split into a pair of disjoint expanding ($+$) and
contracting ($-$) solutions, regarding the value of the scalar field $\phi$ as
an internal emergent physical ``clock''.  \emph{All} classical solutions are
therefore singular either in the ``past'' ($\phi\rightarrow-\infty$) or the
``future'' ($\phi\rightarrow+\infty$).%
\footnote{Actually, if one expresses (\ref{eq:FLRWclasssoln}) in terms of 
$\v$, we see there are \emph{four} distinct solutions, two with $\v>0$ and 
two with $\v<0$.  These solutions are physically degenerate.
} %

Solving the Hamiltonian constraint $H\approx 0$ gives the Friedmann equation
\begin{equation}
\left(\frac{b}{\gamma}\right)^2=\frac{8\pi G}{3}\rho
\label{eq:Friedmann}
\end{equation}
for a flat universe, where the scalar matter energy density on the spatial 
slices $\Sigma$ is given by
\begin{equation}
\rho = \frac{p_{\phi}^2}{2V|_{\phi}^2}\, .
\label{eq:rhodef}
\end{equation}
Here $V|_{\phi}^2$ is the volume at scalar field value $\phi$ given by 
(\ref{eq:FLRWclasssoln}).
 
To make the connection with loop quantum cosmology, we note that in LQC the 
natural phase space variables specify the symmetric connection $c$ and its 
conjugate triad $p$, related to the Ashtekar-Barbero $SU(2)$ connection 
$A^i_a$ and the densitized triad $E^a_i$ by
\begin{equation}
A^i_a = c \oV^{-1/3}\ow^i_a \ ,  \qquad
E^a_i = p \sqrt{\oq}  \oV^{-2/3} \oe^a_i\, .
\label{eq:LQCvardefs}
\end{equation}
Here $\oe^a_i$ and $\ow^i_a$ are the fiducial triad and co-triad respectively,
$\oe^a_i \ow^j_a = \delta^j_i$, in terms of which
$\oq_{ab}=\ow^i_a\ow^j_b\delta_{ij}$.  In these variables the physical volume
of $\cell$ is given by $V=|p|^{3/2}$, and
\begin{equation}
b = \frac{c}{|p|^{1/2}}  \ ,  \qquad
\v = \varepsilon\, \frac{|p|^{3/2}}{2\pi\gamma\lp^2}\, .
\label{eq:LQCbnudefs}
\end{equation}
It's easy to check that  $\{c,p\}= (8\pi G/3)\gamma$.
The variables $c$ and $p$ are related to the original geometrodynamic phase 
space variables by
\begin{eqnarray}
c &=& \varepsilon\, \frac{4\pi G}{3}\, \gamma\, \oV^{-2/3}\, \frac{p_a}{a} \ ,
     \qquad  p = \varepsilon\, \oV^{2/3}a^2
\nonumber\\ 
& = & \varepsilon\, \gamma \oV^{1/3}\, \frac{\dot{a}}{N}\, .
\label{eq:LQCcanonrel}
\end{eqnarray}

\subsection{Quantization of scalar cosmologies}
\label{sec:quantcosmo}

We proceed to describe two \emph{physically inequivalent}\cite{ashsingh11} 
quantizations of this classical model, a conventional, so-called 
``Wheeler-DeWitt'' quantization, and a distinct loop quantum gravity-inspired 
quantization called ``sLQC'' (for ``solvable LQC'').

In order to quantize, we must choose a gauge i.e.\ fix the lapse $N(t)$.  The
classical ``harmonic'' gauge $N(t) = a(t)^3$ simplifies the Hamiltonian
(\ref{eq:FLRWHam}), and it is this choice that leads to the exact solvability
of the loop quantization.  In this gauge the Hamiltonian may be written in the
variables $(b,\v)$ as
\begin{equation}
H  =  \frac{1}{2\oV}\left\{ -3\pi G\hbar^2\, b^2\v^2 + p_{\phi}^2  \right\}.
\label{eq:LQCHam}
\end{equation}
 This is the Hamiltonian we will quantize in the sequel.

It should be noted that in spite of the fact that $-\infty < \v < \infty$, in
both the Wheeler-DeWitt and loop quantizations the $\v \lessgtr 0$ sectors are
both disjoint and physically identical in the absence of fermions, differing
only in triad orientation.  Physical states may be assumed to be symmetric in
$\v$, and we will typically restrict attention to the $\v > 0$ sector.  (See
Ref.~\refcite{ashsingh11} for further discussion of this point.)

\subsubsection{Wheeler-DeWitt quantization}
\label{sec:WdWqc}

The conventional quantization proceeds as usual by defining the conjugate 
operators
\begin{align}
\hat{\v} &= \v    &  
\hat{b} & = 2i\partial_{\v}
\nonumber\\
\hat{\phi} &= \phi    &     
\hat{p}_{\phi} &= -i\hbar\partial_{\phi}
\label{eq:physopdefs}
\end{align}
which satisfy $[\hat{b},\hat{\v}]=2i$ and 
$[\hat{\phi},\hat{p}_{\phi}]=i\hbar$, in accord with the Poisson brackets 
given above.  The Hamiltonian constraint, $H\approx 0$, becomes the 
Wheeler-DeWitt equation $\hat{H}\PsiW =0$,
\begin{eqnarray}
\partial_{\phi}^2\PsiW(\v,\phi)  & = & \k^2\, \frac{1}{\sqrt{|\v|}}\,
   \v\partial_{\v}(\v\partial_{\v}(\sqrt{|\v|}\PsiW(\v,\phi)))
\nonumber\\
 & \equiv & \TW\PsiW(\v,\phi),
\label{eq:WdWEOM}
\end{eqnarray}
with a choice of operator ordering corresponding to the Laplace-Beltrami 
operator of the DeWitt metric on the configuration space $(\v,\phi)$.%
\footnote{This is the choice adopted in Refs.~\refcite{dac13a,CS13a}, and is a
slightly different representation than that employed in Ref.~\refcite{CS10c}.
In contrast to that reference, our states carry an additional factor of
$\sqrt{\lambda/|\v|}$ in order to simplify the form of the inner product.
} %

Group averaging leads to a solvable quantum theory in which physical states 
may be chosen to be ``positive frequency'' solutions to the quantum constraint
\begin{equation}
+\hat{p}_{\phi}\PsiW(\v,\phi) = \hbar \sqrt{\TW}\PsiW(\v,\phi).  
\label{eq:pdef}
\end{equation}
(The positive and negative frequency sectors are disjoint and physically 
equivalent.)  Thus
\begin{equation}
\PsiW(\v,\phi) = \UW(\phi-\phi_0)\,\PsiW(\v,\phi_0),
\label{eq:WdWevolve}
\end{equation}
where the ``propagator'' in internal ``time'' $\phi$ is given by
\begin{equation}
\UW(\phi) = e^{i\sqrt{\TW}\phi} \ .
\label{eq:WdWpropdef}
\end{equation}
The gravitational ``evolution'' operator $\TW$ is positive and self-adjoint in 
the group-averaged, Schr\"{o}dinger-like inner product%
\footnote{It is perhaps worth some emphasis that the form of the inner 
product in both the Wheeler-DeWitt and loop quantizations depends on the 
chosen representation of states.  For example, in some variables the inner 
product assumes a Klein-Gordon type form; see Refs.~\refcite{acs:slqc,CS10c} 
for some discussion and examples.
} %
\begin{equation}
\bracket{\PsiW}{\PhiW} = \int_{-\infty}^{\infty}d\v\, 
\PsiW(\v,\phi)^*\PhiW(\v,\phi).
\label{eq:WdWip}
\end{equation}
Solutions may be conveniently expressed in terms of the eigenfunctions of 
$\TW$,
\begin{equation}
\eW_k(\v) = \frac{1}{\sqrt{4\pi|\v|}}\,e^{ik\ln\left|\frac{\v}{\lambda}\right|},
\label{eq:WdWedef}
\end{equation}
where
\begin{equation}
\TW\eW_k = \omega_k^2\, \eW_k.
\label{eq:WdWTdef}
\end{equation}
Here $\omega_k = \k\, |k|$, and we choose 
$\lambda =  \sqrt{\Delta} \cdot \lp =\sqrt{4\sqrt{3}\pi\gamma} \cdot \lp$. %
($\lambda^2$ is the so-called ``area gap'' of loop quantum gravity.)  This
specific choice for the constant $\lambda$ -- necessary for dimensional
reasons -- plays no physical role in the Wheeler-DeWitt theory, but is a
convenient normalization for comparison with LQC. In terms of the $\eW_k(\v)$,
physical states may be written
\begin{eqnarray}
\PsiW(\v,\phi) & = & \int_{-\infty}^{\infty}dk\, 
\PsiWk(k)\,\eW_k(\v)\,e^{i\omega_k(\phi-\phi_0)}
   \nonumber \\
 & = &  \PsiWR(\v,\phi) + \PsiWL(\v,\phi),
\label{eq:WdWQWFdef}
\end{eqnarray}
where
\begin{eqnarray}
\PsiWR(\v,\phi) & = & \frac{1}{\sqrt{4\pi|\v|}}\, \int_{-\infty}^{0}dk\, 
            \PsiWk(k)\,e^{ik[\ln\left|\frac{\v}{\lambda}\right|-\k(\phi-\phi_0)]}
\nonumber\\
\PsiWL(\v,\phi) & = & \frac{1}{\sqrt{4\pi|\v|}}\, \int_{0}^{\infty}dk\, 
            \PsiWk(k)\,e^{ik[\ln\left|\frac{\v}{\lambda}\right|+\k(\phi-\phi_0)]}.
\label{eq:WdWPsiLRdef}
\end{eqnarray}
These ``right-'' and ``left-''moving (in a plot of $\phi$ vs.\ $\v$) states 
clearly correspond to the expanding and contracting branches of the classical 
solution.  In the quantum theory, the $R$ and $L$ sectors are orthogonal and 
a-priori independent of one another.   A state may be purely $R$- (or $L$-) 
moving, or a superposition of the two.

For considerable further detail concerning the quantization and 
references to the earlier literature see Ref.~\refcite{ashsingh11}.

\subsubsection{Observables}
\label{sec:observe}

The physical (``Dirac'') observables must be represented by operators that
commute with the constraint $\TW$.  The scalar momentum $\hat{p}_{\phi}$
clearly commutes with $\sqrt{\TW}$ via (\ref{eq:WdWEOM}) and is therefore a
constant of the motion as in the classical theory.  The volume $\hat{\v}$ is
not.  However, the corresponding ``relational'' observable
$\hat{\v}|_{\phi^*}$ that gives the value of the volume at a fixed value
$\phi^*$ of the scalar field is.  The ``Heisenberg-picture'' operator
$\hat{\v}|_{\phi^*}(\phi)$ that acts on states at $\phi$ is given by
\begin{equation}
\hat{\v}|_{\phi^*}(\phi) = U(\phi^*-\phi)^{\dagger}\hat{\v} U(\phi^*-\phi),
\label{eq:nurelopdef}
\end{equation}
where $U(\phi)$ is given by (\ref{eq:WdWpropdef}).  Thus, for example, the 
physical volume of the fiducial cell $\cell$ at $\phi^*$ is given by the 
operator $\hat{\v}|_{\phi^*}(\phi)$ whose action is
\begin{equation}
\hat{\v}|_{\phi^*}(\phi)\, \Psi(\v,\phi) = 2\pi\gamma\lp^2\, 
  e^{i\sqrt{\TW}(\phi-\phi^*)} |\v|\,\Psi(\v,\phi^*).
\label{eq:nurelopaction}
\end{equation}

FLRW universes are, of course, classically singular -- every classical
solution (\ref{eq:FLRWclasssoln}) begins or ends in a zero-volume singularity
with diverging energy density.  In Ref.~\refcite{acs:slqc} it is shown that
the same is true of the Wheeler-DeWitt quantum theory in the sense that the
expectation value of the volume observable vanishes for generic right-moving
(expanding) states as $\phi\rightarrow -\infty$, and for generic left-moving
states as $\phi\rightarrow +\infty$.  (This is actually evident from Eq.\
(\ref{eq:WdWPsiLRdef}) by simply invoking the Riemann-Lebesgue lemma.)  Since
$\hat{p}_{\phi}$ is a constant of the motion this implies the matter energy
density (\ref{eq:rhodef}) diverges in those limits for generic states.  By
constrast, they showed that in sLQC the (expectation value of the) volume for
generic states remains bounded away from zero, and the (expectation value of
the) matter energy density is bounded above for generic states.%
\footnote{Prior work\cite{aps,aps:improved} employing lapse $N=1$ had shown 
this numerically for sharply peaked states (later confirmed for generic 
states.\cite{dgs14a,dgms14a})  The exact solvability of the theory in the 
harmonic gauge makes an analytic result possible in sLQC.
} %
Here we will sharpen these results and demonstrate the singularity of the
Wheeler-DeWitt quantization and concomitant finiteness of LQC using the
consistent histories framework for quantum theory.

\subsubsection{Loop quantization}
\label{sec:loopqc}

The loop quantization proceeds by quantizing the Hamiltonian (\ref{eq:LQCHam})
on a different kinematical Hilbert space.  (For details and references to the
earlier literature see Ref.~\refcite{ashsingh11}.)  If $\ket{\v}$ denotes the
eigenstates of the multiplicative volume operator $\hat{\v}$, in this
inequivalent quantization one finds that $\widehat{\exp(i\lambda
b)}\ket{\v}=\ket{\v-2\lambda}$ acts as a translation.  Quantizing the
Hamiltonian constraint once again leads to
\begin{equation}
\hat{p}_{\phi}^2\,\Psi(\v,\phi) = \hbar^2\, \Theta\, \Psi(\v,\phi),
\label{eq:LQCEOM}
\end{equation}
but now $\Theta$ is a second-order difference operator
\begin{eqnarray}
\Theta\, \Psi(\v,\phi) & = &
-\frac{3\pi G}{4\lambda^2} \left\{
\sqrt{|\v(\v+4\lambda)|} |\v+2\lambda| \Psi(\v+4\lambda,\phi)
- 2 \v^2 \Psi(\v,\phi)  \right. \nonumber\\
  & \qquad &  \qquad \qquad \qquad \qquad 
 \left. + \sqrt{|\v(\v-4\lambda)|} |\v-2\lambda| \Psi(\v-4\lambda,\phi) \right\}. 
\label{eq:LQCTopdef}
\end{eqnarray}
Solutions to the full quantum constraint $\hat{C}=-[\partial_{\phi}^2+\Theta]$ 
therefore decompose into disjoint sets of solutions on the 
$\epsilon$-lattices given by $\v=4\lambda n + \epsilon$, where $\epsilon \in 
[0,4\lambda)$.  Without loss of generality we will restrict attention to the 
sector of the theory on the $\epsilon=0$ latice.  Thus, volume in this model 
is discrete,
\begin{equation}
  \v = 4\lambda n  \ , \quad n \in \Int.
\label{eq:nulattice}
\end{equation}
Group averaging leads to the physical inner product 
\begin{equation}
\bracket{\Psi}{\Phi} = \sum_{\v=4\lambda n} \Psi(\v,\phi_0)^*\Phi(\v,\phi_0)
\label{eq:LQCipdef}
\end{equation}
at some fiducial (but physically irrelevant) $\phi_0$.  Once again the theory
splits into disjoint, physically degenerate positive- and negative-frequency
sectors, and as before we restrict our attention to the positive frequency
sector
\begin{equation}
-i\partial_{\phi} \Psi(\v,\phi) = \sqrt{\Theta}\, \Psi(\v,\phi).
\label{eq:LQCEOMroot}
\end{equation}
Also as in the Wheeler-DeWitt theory, states in sLQC may be ``propagated'' by
$U(\phi)=\exp(i\sqrt{\Theta}\phi)$.  Relational Dirac observables in sLQC are
then defined in precisely the same manner as in the Wheeler-DeWitt theory.

\begin{figure}[!htbp]   
\begin{center}
\includegraphics[width=1.0\textwidth]{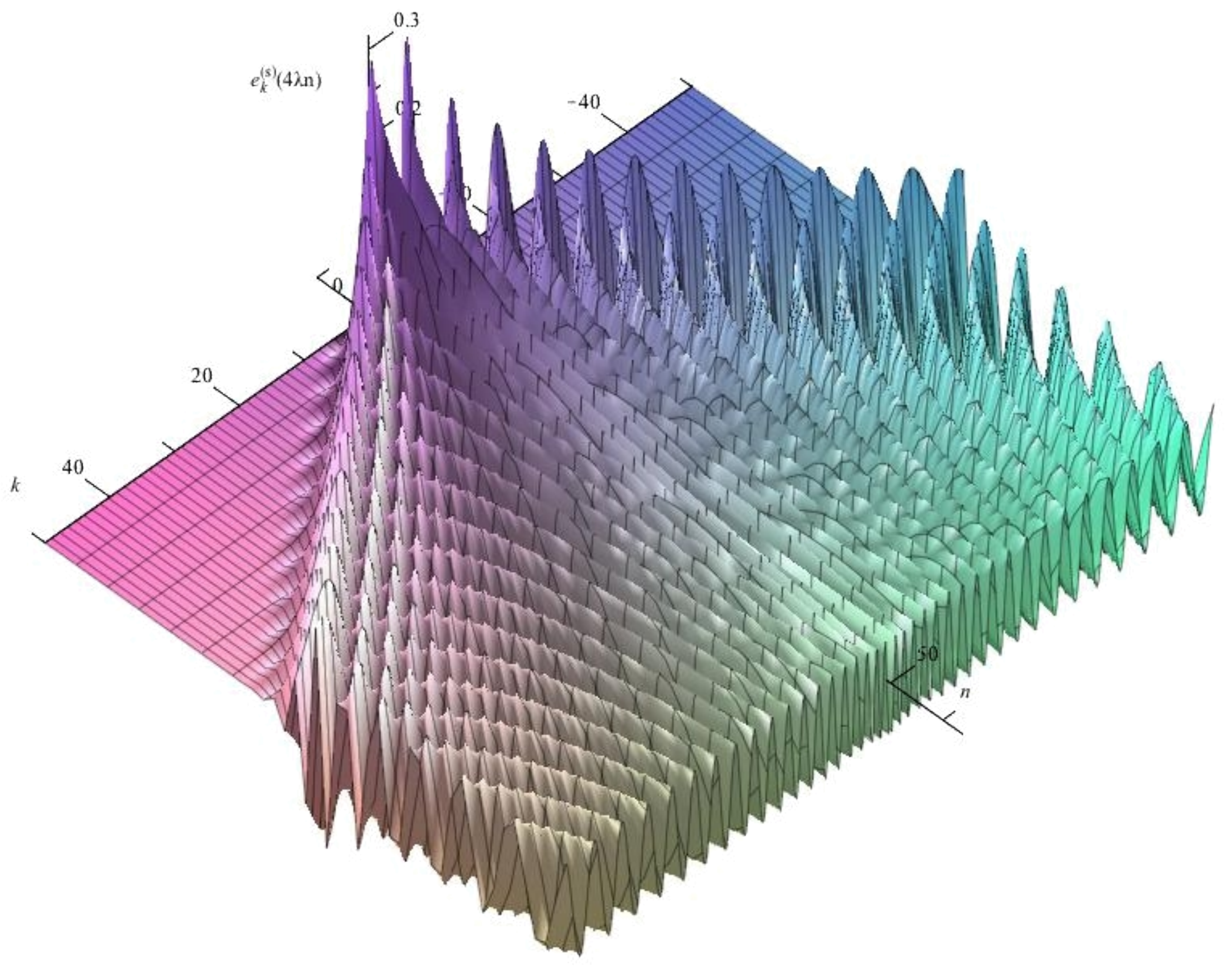}
\end{center}
\vspace{-15pt}
\caption{Plot of the gravitational functions $\eS_k(\v=4\lambda n)$ in the
$(k,n)$ plane.  The volume variable $\v=4\lambda n$ is fundamentally discrete;
the values of the eigenfunctions are plotted as continuous in both variables
$k$ and $n$ for reasons of visual clarity only.  
The exponential ultraviolet cutoff along the lines $|k|=2|n|=|\v/2\lambda|$ is
clearly evident.  The eigenfunctions $\eS_k(\v)$ may therefore
be regarded to an excellent approximation as having support only in the
``wedge'' $|k|\lesssim 2|n|$.  It is this feature of the eigenfunctions that
is ultimately responsible for the existence of a universal upper bound to the
matter density.
}%
\label{fig:esWedge}%
\end{figure}

Similarly, quantum states may be represented in terms of the symmetric 
eigenfunctions $\eS_k(\v)$ of $\Theta$,
\begin{equation}
\Theta \eS_k(\v) = \omega_k^2\, \eS_k (\v) \ ,
\label{eq:LQCedef}
\end{equation}
as
\begin{equation}
\Psi(\v,\phi)  =  \int_{-\infty}^{\infty}dk\, 
\Psik(k)\,\eS_k(\v)\,e^{i\omega_k(\phi-\phi_0)}.
\label{eq:LQCQWFdef}
\end{equation}
Explicit expressions for the eigenfunctions $\eS_k(\v)$, based on the earlier
work of Refs.~\refcite{ach09,ach10a,ach10b}, are given in \refcite{dac13a}.
(See Eq.  (3.14) of that reference.)  Their salient properties are
\begin{description}[leftmargin=0cm]
\item[(i)] They are oscillatory  functions of both $k$ and $\v$, oscillating 
increasingly rapidly in $k$ as $\v$ increases.  They are symmetric in both 
$k$ and $\v$.

\item[(ii)] At large volume (specifically, for $|\v| \gg 2\lambda\, |k|$), 
they rapidly approach a specific linear combination of corresponding 
Wheeler-DeWitt eigenfunctions, namely
\begin{eqnarray}
\eS_k(\v) & \simeq & \sqrt{\frac{2\lambda}{\pi |\v|}}\,
    \cos(|k|\ln\left|\frac{\v}{\lambda}\right|+\alpha(|k|))   
       \qquad   |\v| \gg 2\lambda |k|
\nonumber\\
 & = & \sqrt{2\lambda}\,  
    \left(\eW_{+|k|}(\v)e^{+i\alpha(|k|)}  + \eW_{-|k|}(\v)e^{-i\alpha(|k|)} \right),
\label{eq:escasslim}
\end{eqnarray}
where $\alpha(k) = k\ln(1-\ln k) + \frac{\pi}{4}$.

\item[(iii)]  Regarded as functions of $k$, the $\eS_k(\v)$ exhibit a sharp 
exponential ultraviolet cutoff that sets in when $2\lambda |k| \approx 
|\v|$.  In other words, the eigenfunctions have support only in the wedge
\begin{equation}
  |k| \lesssim \frac{|\v|}{2\lambda}.
\label{eq:kwedge}
\end{equation}

\end{description}
All of these features are evident in Fig.\ \ref{fig:esWedge}. 

Property (ii) is worthy of note because it directly implies that \emph{all} 
states in LQC approach a specific, \emph{symmetric} linear combination of 
expanding and contracting Wheeler-DeWitt states
\begin{equation}
\Psi(\v,\phi) \approx \PsiWR(\v,\phi) + \PsiWL(\v,\phi)  
\qquad 
\mbox{\scriptsize $
\begin{pmatrix}
\text{large}\\
\text{volume}  
\end{pmatrix}
$}     
\label{eq:LQCLRlim}
\end{equation}
at large volume.\cite{dac13a} (See Refs.~\refcite{dac13a,CS13a} for
considerable further detail and discussion of this point.  By contrast, recall
in the Wheeler-DeWitt theory the expanding and contracting branches of a
generic state are completely independent of one another.)  This is one
signature of the quantum ``bounce'' characteristic of loop quantum states.
Indeed, it is well known that in LQC quasiclassical states (as described in
the sequel) remain peaked on a solution of the quantum-corrected Friedmann
equation\cite{singh06a,taveras08a}
\begin{equation}
\left(\frac{\dot{a}}{a}\right)^2 = \frac{8\pi G}{3}\rho\left(1-\frac{\rho}{\rhom}\right)
\label{eq:LQCFriedmann}
\end{equation}
for all values of the volume, ``bouncing'' from one large volume state to
another rather than being swallowed by the singularity as in the classical or
Wheeler-DeWitt theory; see Fig.~\ref{fig:msstraj}.  Here $\rhom$ is the
maximum matter energy density defined in (\ref{eq:rhocritdef}) below.
Equation (\ref{eq:LQCLRlim}) is an indication of the remarkable fact,
demonstrated in further detail in the sequel, that \emph{generic} quantum
states in sLQC are
nonsingular\cite{aps,aps:improved,acs:slqc,ashsingh11,dac13a,CS13a,dgs14a,dgms14a}
-- \emph{all} states ``bounce'', not merely quasiclassical ones.

In this regard property (iii) of the eigenfunctions is especially interesting,
because it is directly responsible for the non-singular nature of sLQC. It is
a manifestation of the quantum gravitational repulsion that appears at the
Planck scale in LQG. In fact, one may argue heuristically that the matter
energy density remains bounded in LQC by
\begin{equation}
\rhom = \frac{\sqrt{3}}{32\pi^2\gamma^3}\, \rho_p,
\label{eq:rhocritdef}
\end{equation}
where $\rho_p=1/G\lp^2$ is the Planck density.   The argument proceeds as 
follows.  The matter density is given classically by (\ref{eq:rhodef}).  It 
is argued in Ref.~\refcite{acs:slqc}  
that correspondingly
\begin{equation}
\expct{\rho|_{\phi}} = 
  \frac{1}{2}\, \frac{\expct{p_{\phi}^2}}{\expct{\hat{V}|_{\phi}^2}}
\label{eq:rhoopdef}
\end{equation}
in the quantum theory.  (Variations on this definition have also been
discussed.\cite{acs:slqc,dac13a})  Now, the eigenvalues of $\hat{p}_{\phi}$
are $\hbar\k k$, and the UV cutoff (\ref{eq:kwedge}) on the eigenfunctions
$\eS_k(\v)$ requires $|k| \lesssim |\v|/2\lambda$, so that with
$\hat{V}=2\pi\lp^2\gamma|\hat{\v}|$,
\begin{eqnarray}
\expct{\rho|_{\phi}} & \sim & \frac{1}{2}\, \frac{(\hbar\k |k|)^2}{(2\pi\lp^2 
\gamma |\v|)^2}
\nonumber\\
 & \lesssim & \frac{1}{2} \left(\frac{\hbar\k}{2\lambda}\right)\left( 
\frac{|\v|}{2\pi\gamma\lp^2 |\v|}\right)^2,
\label{eq:rholim}
\end{eqnarray}
which gives precisely the bound (\ref{eq:rhocritdef}).  Thus, the linear 
scaling of the UV cutoff on the eigenfunctions with volume leads to a 
universal upper bound on the expectation value (hence spectrum) of the matter 
energy density.
It is satisfying that this simple heuristic argument based on the UV cutoff
(\ref{eq:kwedge}) reproduces precisely the value of the bound on the density
found by a careful analytical argument in Ref.~\refcite{acs:slqc}.  A quite
different analytical argument may be found in Ref.~\refcite{dac13a}.

%

\section{Consistent Histories Formulation of Canonical Quantum Cosmology}
\label{sec:chqc}

Both the Wheeler-DeWitt quantization and sLQC are canonical quantum theories, 
in the sense that they describe quantum universes by states in well-defined 
Hilbert spaces, with physical observables represented (Dirac) operators on 
those Hilbert spaces.  This means that each may be given a consistent 
histories formulation that closely resembles that of non-relativistic 
quantum theory.

Indeed, the presence in these models of a variable that behaves as an emergent
internal matter ``clock'', the scalar field $\phi$ (via
$\ket{\Psi(\phi)}=U(\phi)\ket{\Psi}$), only encourages that association.  We
emphasize, however, that that association is not \emph{necessary} for the
formulation of the consistent histories framework.  What matters is that
well-defined quantum states and observable operators are present which permit
the definition of class operators and branch wave functions out of which a
decoherence functional may be constructed.  The decoherence functional can
then be employed to make predictions concerning patterns of \emph{correlation}
between any of the observable quantities, whether or not they are necessarily
ordered in some ``time'' variable.  The existence of an internal time,
however, does supply a convenient context for the physical interpretation of
the resulting theory.

\subsection{Decoherence functional}
\label{sec:df}

In order to define the decoherence functional for these theories we must 
first construct class operators and branch wave functions for quantum 
histories defined by histories of values of the theory's Dirac observables.  
The constructions are identical in each theory so long as they are expressed 
in terms of the gravitational constraint operator $\Theta$ and its 
eigenfunctions.  The differing physical predictions then arise because of the 
sharply distinct behavior of these objects in the two theories.  We therefore 
proceed to formulate the decoherence functional in terms of $\Theta$ and its 
eigenfunctions alone, and apply the construction to contrast the physical 
predictions of the two theories in the sequel.

In this simple FLRW model our operators are 
$\{\hat{\phi},\hat{p}_{\phi},\hat{b}\, (\text{or }\widehat{\exp(i\lambda 
b)}),\hat{\v}\}$ with corresponding relational Dirac observables given by the 
equivalents of (\ref{eq:nurelopdef}).  Given its relevance to the question of 
the singularity of the universe we will concentrate most of our attention on 
$\hat{\v}$, but parallel constructions are available for any observable.

The definition of class operators is based on the spectral decomposition  of 
observables.  In the case of the volume operator
\begin{eqnarray}
\hat{\v}^{\sWdW} & = & 
\int_{-\infty}^{\infty}d\v'\, \v'\, \ketbra{\v'}{\v'}
\nonumber\\
 & = & \int_{-\infty}^{\infty}d\v'\, \v'\, P^{\v}_{\v'}\ ,
\label{eq:nuWdWopspectral}
\end{eqnarray}
with 
\begin{eqnarray}
\bracket{\v}{\v'} & = & \delta(\v-\v') 
\nonumber\\
e^{\sWdW}_k(\v) & = & \bracket{\v}{k^{\sWdW}}
\label{eq:nuketdefwdw}
\end{eqnarray}
and $P^{\v}_{\v'}$ a volume projection on the eigenket $\ket{\v'}$.  (We 
work with a normalization appropriate to the full range $-\infty < \v < 
\infty$ and take all states to be symmetric in $\v$.)  By contrast, in sLQC 
volume is discrete:
\begin{eqnarray}
\hat{\v} & = & \sum_{\v=4\lambda n} \v\, \ketbra{\v}{\v}
\nonumber\\
 & = & \sum_{\v'=4\lambda n'} \v'\, P^{\v}_{\v'}.
\label{eq:nuopspectral}
\end{eqnarray}
and 
\begin{eqnarray}
\bracket{\v}{\v'} & = & \delta_{n,n'}
\nonumber\\
\eS_k(\v) & = & \bracket{\v}{k^{{\scriptscriptstyle (s)}}}\ .
\label{eq:nuketdefslqc}
\end{eqnarray}
In the sequel $\hat{\v}$ will refer to whichever theory is appropriate in 
context, with integrals in the Wheeler-DeWitt theory and sums in sLQC.  For 
example, the projection onto a range of values $\Delta\v$ will be one of
\begin{equation}
P^{\v}_{\Delta\v} =
\begin{cases}
\int_{\v\in\Delta\v}d\v\, \ketbra{\v}{\v} & \text{Wheeler-DeWitt} \\
\sum_{\v\in\Delta\v}\ketbra{\v}{\v} & \text{slQC}  \ .
\end{cases}
\label{eq:nuprojdef}
\end{equation}
``Heisenberg'' projections may be defined via the propagator 
$U(\phi)=\exp(i\sqrt{\Theta}\phi)$ as
\begin{equation}
P^{\alpha}_{\Delta a^{\alpha}_{k}}(\phi) =   U(\phi-\phi_0)^{\dagger}
P^{\alpha}_{\Delta a^{\alpha}_{k}} U(\phi-\phi_0),
\label{eq:heisprojdef}
\end{equation}
where $\phi_0$ is a fiducial (but physically irrelevant) value of the scalar
field at which the quantum state is defined.  Class operators may then be
defined as in Eqs.~(\ref{eq:classopdef-fg}-\ref{eq:classopdef-cg}).  For
example, the class operator for the relational history in which the volume is
in $\Delta\v$ when the scalar field has value $\phi^*$ is
\begin{equation}
C_{\Delta\v|_{\phi^*}} =  U(\phi^*-\phi_0)^{\dagger}
P^{\v}_{\Delta\v} U(\phi^*-\phi_0).
\label{eq:nuclassopdef}
\end{equation}
It turns out such class operators offer an illuminating perspective on 
relational observables such as $\v|_{\phi^*}(\phi)$, a point to which we 
shall return later.   Similarly, the class operator describing a 
coarse-grained trajectory $\v(\phi)$, for which the volume is in $\Delta\v_1$ 
at scalar field value $\phi_1$, $\Delta\v_2$ at $\phi_2$, and so in, is
\begin{equation}
C_{\Delta\v_1|_{\phi_1};\Delta\v_2|_{\phi_2};\cdots;\Delta\v_n|_{\phi_n}}
= \Projsupb{\v}{\Delta \v_1}(\phi_1) \Projsupb{\v}{\Delta \v_2}(\phi_2)
      \cdots \Projsupb{\v}{\Delta \v_n}(\phi_n).
\label{eq:nutrajclassopdef}
\end{equation}
Branch wave functions corresponding to a history $h$ are then defined as in 
(\ref{eq:bwfdef}) by
\begin{equation}
\ket{\Psi_h(\phi)} = U(\phi-\phi_0) C_h^{\dagger}\ket{\Psi},
\label{eq:qcbwfdef}
\end{equation}
which is everywhere a solution of the theory's Wheeler-DeWitt equation (i.e.\ 
annihilated by the quantum constraint.)  The decoherence functional in the 
case of a pure ``initial'' state defined on a minisuperspace surface of 
constant $\phi$%
\footnote{See \refcite{CS10c} for brief discussion of this point.
} %
may then be defined simply as 
\begin{equation}
d(h,h') = \bracket{\Psi_{h'}}{\Psi_h},
\label{eq:qcdfdef}
\end{equation}
using the group-averaged inner product on states.  Consistent or decoherent
families of histories then satisfy (\ref{eq:dfndmtl}) on all pairs of
histories $h$ in some exclusive, exhaustive family $\{h\}$.  As discussed
above, the decoherence functional provides an objective, observer-independent
measure of the interference among histories in the family.  When that
interference vanishes, each history in the family may be assigned the
probability $p(h)=d(h,h)$.  Otherwise, quantum mechanics says that the
physical question posed by the family of histories simply has no sensible
answer within quantum theory.

Let us see how this is accomplished in a series of examples in keeping with 
the plan laid out in section \ref{sec:prediction}.

\subsection{Consistent histories in quantum cosmology: other models}
\label{sec:other}

sLQC has also been used as the basis for a spin-foam-like ``path integral''
formulation of loop quantum cosmology.\cite{ach09,ach10a,ach10b,ashsingh11}
(See also related work.\cite{hrvwe11}) As remarked at the end of section
\ref{sec:gqt}, a consistent histories formulation may also be given for
theories defined via path-integrals, as has been done (for example) for
Bianchi IX cosmologies and other other
models.\cite{hartle91a,dac07,CH04,halliwell99,hallithor01,hallithor02,halliwall06,halliwell09}
These constructions serve as the template for a consistent histories
formulation of spin foam loop quantum cosmology\cite{CS16a,CS16b,CS16c} and of
spin foam loop quantum gravity (as in Ref.~\refcite{schroer13a}, which is
directly modeled on Refs.~\refcite{hartle91a,lesH,CH04}.)  We do not, however,
have the space to describe these path-integral consistent histories theories
here.

It should also be mentioned that the physical predictions of the consistent
histories formulation have been compared to a de Broglie-Bohm quantization of
FLRW that closely parallels the work reviewed herein.\cite{fpps12,pf13a}
See, however, Ref.~\refcite{CS13a} for some pertinent brief discussion.

\section{Applications}
\label{sec:app}


\subsection{Probabilities, histories, and relational observables}
\label{sec:relnlobs}

Relational (``Dirac'') observables are the physical quantities about which
theories with constraints make predictions, and (by definition) must commute
with those constraints.  It might have been thought that the class operators
for volume given in (\ref{eq:nuclassopdef}-\ref{eq:nutrajclassopdef}) should
have been constructed directly from the Dirac observable
(\ref{eq:nurelopdef}).  While that indeed is an option, an alternative point
of view is that histories provide a natural framework within which to
understand the \emph{emergence} of
such relational observables 
in theories with an emergent ``time'' evolution.\cite{CS10c} %

To see how they arise naturally in a histories framework, consider a 
self-adjoint operator $\hat{A}$ that does not commute with the constraint -- 
for example, $\hat{\v}$, in the models we have been discussing -- with 
spectral decomposition (assuming for definiteness that $\hat{A}$ has a 
purely discrete spectrum)  
\begin{equation}
\hat{A} = \sum_a a\, P_a,
\label{eq:Aspectral}
\end{equation}
where $P_a=\ketbra{a}{a}$.  %
The class operator corresponding to histories in which $\hat{A}$ has values in
one of the complete set of disjoint of intervals of eigenvalues $\{\Delta a\}$
at $\phi=\phi^*$ are
\begin{equation}
C_{\Delta a|_{\phi^*}}  =  
  U(\phi^*-\phi_0)^{\dagger} \Projsupb{A}{\Delta a} U(\phi^*-\phi_0),
\label{eq:classopA}
\end{equation}
with the corresponding branch wave functions defined as in 
(\ref{eq:qcbwfdef}).  Because the class operators are simply one ``time'' 
($\phi$) projections, the branch wave functions \emph{always} decohere -- the 
family is consistent:
\begin{eqnarray}\label{eq:drlnl}
d(\Delta a,\Delta a') & = & 
\bracket{\Psi_{\Delta a'|_{\phi^*}}}{\Psi_{\Delta a|_{\phi^*}}}
\nonumber\\
& = & 
\melt{\Psi}{C_{\Delta a'|_{\phi^*}}C_{\Delta a|_{\phi^*}}^{\dagger}}{\Psi}
\nonumber\\
& = & 
\melt{\Psi}{U(\phi^*-\phi_0)^{\dagger} \Projsupb{A}{\Delta a'} U(\phi^*-\phi_0)
   U(\phi^*-\phi_0)^{\dagger} \Projsupb{A}{\Delta a} U(\phi^*-\phi_0)}{\Psi}
\nonumber\\
& = & 
\melt{\Psi(\phi^*)}{\Projsupb{A}{\Delta a'}\Projsupb{A}{\Delta a}}{\Psi(\phi^*)}
\nonumber\\
& = & 
\melt{\Psi(\phi^*)}{\Projsupb{A}{\Delta a}}{\Psi(\phi^*)}\,\delta_{\Delta a',\Delta a}
\nonumber\\
& = &  p_{\Delta a}(\phi^*)\,\delta_{\Delta a',\Delta a}\ ,
\end{eqnarray}
where $p_{\Delta a}(\phi^*)$ is the probability that $a\in\Delta a$ when
$\phi=\phi^*$.  For example, since we have assumed $\hat{A}$ has a discrete
spectrum if we set $\Delta a = \{a\}$, a single eigenvalue, then
\begin{equation}
p_a(\phi^*) = |\bracket{a}{\Psi(\phi^*)}|^2
\label{eq:Aprob}
\end{equation}
as one might expect,%
\footnote{Precisely because of this expectation, it is crucial to emphasize
that the simple form of this result, which we have derived from the
decoherence functional, is directly connected with the simple form of the
Schr\"{o}dinger-like form of the inner product in this representation.  As
noted above, 
in other representations the inner product can take on e.g.\ a Klein-Gordon 
type form, and therefore the formula for the probability does as well.  This 
illustrates the importance of placing quantum prediction in a coherent, 
self-consistent frame.\cite{CH04,halliwell91:qcbu}
} %
with a similarly expected expression if $\hat{A}$ has a continuous spectrum; 
see Ref.~\refcite{CS10c} for details in that case.

To see the connection with relational observables, let us calculate the 
average value of $\hat{A}$ at $\phi^*$:
\begin{eqnarray}
\expct{\hat{A}}|_{\phi^*} & = & \sum a\, p_a(\phi^*)
\nonumber\\
 & = & \melt{\Psi(\phi^*)}{\sum_a a\, P^A_a}{\Psi(\phi^*)}
\nonumber\\
 & = & \melt{\Psi}{U(\phi^*-\phi_0)^{\dagger}\hat{A}U(\phi^*-\phi_0)}{\Psi}
\nonumber\\
 & = & \melt{\Psi}{\hat{A}|_{\phi^*}}{\Psi},
\label{eq:Aexpct}
\end{eqnarray}
the expectation value of the relational observable $\hat{A}|_{\phi^*}$
corresponding to $\hat{A}$ in the state $\ket{\Psi}$.  Probabilities for
histories of values of $\hat{A}$ (which does not commute with the constraint)
are naturally expressed in terms of the corresponding relational observable
$\hat{A}|_{\phi^*}$ (which does).  Had we not known about Dirac observables 
the histories formulation would have led us to consider them.

\subsection{Scalar momentum}
\label{sec:scalarmom}

The scalar field momentum $p_{\phi}$ is a constant of the motion in the 
classical theory, where $\{p_{\phi},H\}=0$, and similarly in the quantum 
theory, $[\hat{p}_{\phi},\Theta]=0$.  How does this constancy manifest in the 
consistent histories formulation?

If we are only interested in the distribution of probabilities for values of 
$\hat{p}_{\phi}$ at a single $\phi=\phi^*$, we might construct the relational 
observable $\hat{p}_{\phi}|_{\phi^*}$.  However, because $\hat{p}_{\phi}$ 
commutes with the constraint, $\hat{p}_{\phi}|_{\phi^*}=\hat{p}_{\phi}$, and 
correspondingly the probability $p_{\Delta p_{\phi}}(\phi^*)$ 
for $\hat{p}_{\phi}$ to have values in $\Delta p_{\phi}$ at $\phi^*$ is 
independent of $\phi^*$, in keeping with $p_{\phi}$ being a constant of the 
motion.

Similarly, to the question ``what is the likelihood $p_{\phi}$ is in $\Delta 
p_{\phi;1}$ at $\phi_1$,  $\Delta p_{\phi;2}$ at $\phi_2$, \ldots (etc.)'' 
corresponds the class operator
\begin{equation}
C_{\Delta p_{\phi;1}|_{\phi_1};\Delta p_{\phi;2}|_{\phi_2};\cdots;\Delta p_{\phi;n}|_{\phi_n}}  
= \Projsupb{p_{\phi}}{\Delta p_{\phi;1}}(\phi_1) \Projsupb{p_{\phi}}{\Delta p_{\phi;2}}(\phi_2) 
      \cdots \Projsupb{p_{\phi}}{\Delta p_{\phi;n}}(\phi_n),
\label{eq:pphiclassopdef}
\end{equation}
Since $P^{p_{\phi}}_{\Delta p_{\phi}}(\phi) =P^{p_{\phi}}_{\Delta 
p_{\phi}}$, all such class operators are zero unless all of the 
intervals  $\Delta p_{\phi;i}$ are equal, in which case each such class 
operator is the simple projection $P^{p_{\phi}}_{\Delta p_{\phi;i}}$.  The 
family of all such class operators clearly decoheres for any initial state, 
the corresponding probabilities $p_{\Delta p_{\phi;i}}$ all constants (i.e.\ 
independent of $\phi$).  This is the meaning of the statement that $p_{\phi}$ 
is a constant of the motion in the quantum theory.
(Constants of motion in generalized consistent histories quantum theory are 
discussed further in Ref.~\refcite{hartlemarolf97}.)

\subsection{Volume at a single value of $\phi$}
\label{sec:volphi}

\begin{figure}[hbt!]
\begin{center}
\includegraphics[width=0.80\textwidth]{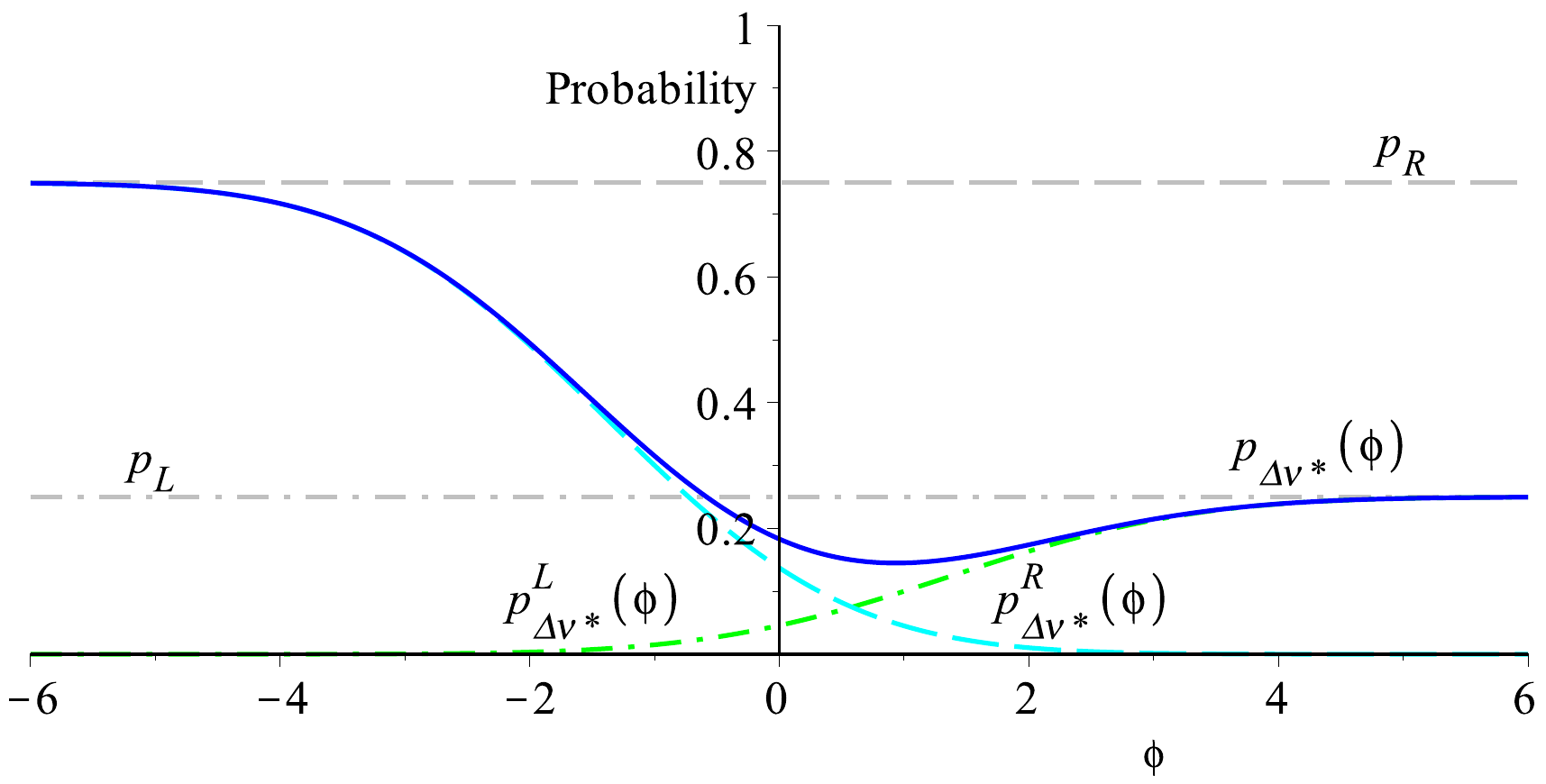}
\end{center}
\vspace{-15pt}
\caption{ The behavior of $p_{\Delta\nu^*}(\phi)$, the probability that the
quantum universe will be found in the interval $\Delta\nu^* =[0,\nu^*]$
(\emph{i.e.\ }at small volume) for a sample superposition of expanding $(R)$
and contracting $(L)$ semiclassical states both peaked at large volume near
$\phi=0$.  $p_L$ and $p_R$ give the relative ``amount'' of each component in
the superposition, so that $p_L + p_R =1$ (cf.~Eq.~(\ref{eq:psiLRcatdef}).)
This plot may appear to imply the possibility of a ``quantum bounce'', since
at any given $\phi$ there is a non-zero probability that the universe may be
found with volume $\nu > \nu^*$.  A more careful consistent histories analysis
shows that this possibility is not realized: the probability that the universe
has large volume in \emph{both} the ``past'' and
``future'' is zero.\cite{CS10b,CS10c}
} %
\label{fig:pvolplot}
\end{figure}

Classically, one may express the volume $\v$ of the fiducial cell as a
function $\v(\phi)$ as in (\ref{eq:FLRWclasssoln}).  It might seem natural
then to ask the quantum question ``what is the probability the volume $\v$ is
in the interval $\Delta\v$ at scalar field value $\phi^*$?''  The class
operator corresponding to this question was given in (\ref{eq:nuclassopdef}).
According to the calculation of (\ref{eq:Aprob}) these single-$\phi$ histories
decohere with corresponding probabilities
\begin{equation}
p_{\Delta\v}(\phi^*) = 
   \bracket{\Psi_{\Delta\v|_{\phi^*}}}{\Psi_{\Delta\v|_{\phi^*}}}.
\label{eq:nuprobdef}
\end{equation}
In the Wheeler-DeWitt case this is, explicitly,
\begin{equation}
p^{\sWdW}_{\Delta\v}(\phi^*) = \int_{\Delta\v}d\v\, 
|\psi^{\sWdW}(\v,\phi^*)|^2.
\label{eq:nuprobWdW}
\end{equation}
This probability is calculated explicitly for semi-classical Wheeler-DeWitt
states (i.e.\ states peaked on a particular classical trajectory) in
Ref.~\refcite{CS10c}.  In the case of a superposition of right- (expanding)
and left- (contracting) moving states (as in (\ref{eq:psiLRcatdef})), an
example of the result is shown in Figure \ref{fig:pvolplot}.\cite{CS10b,CS10c}
In loop quantum cosmology the equivalent expression is
\begin{equation}
p^{\sLQC}_{\Delta\v}(\phi^*) = \sum_{\v\in\Delta\v} |\psi(\v,\phi^*)|^2.
\label{eq:nuprobLQC}
\end{equation}
This is again calculated for the case of ``quasiclassical'' loop quantum states
that approach a symmetric superposition of semiclassical Wheeler-DeWitt states
at large volume in Ref.~\refcite{CS13a}. 
(For discussion of the usage of the term ``quasiclassical'' in this context 
see Sec.~\ref{sec:quasiclassical}.)

\subsection{Cosmological trajectories}
\label{sec:traj}

\begin{figure}[hbtp!]
\begin{center}
\subfloat[Coarse-grained trajectories]{
\includegraphics[width=0.485\textwidth]{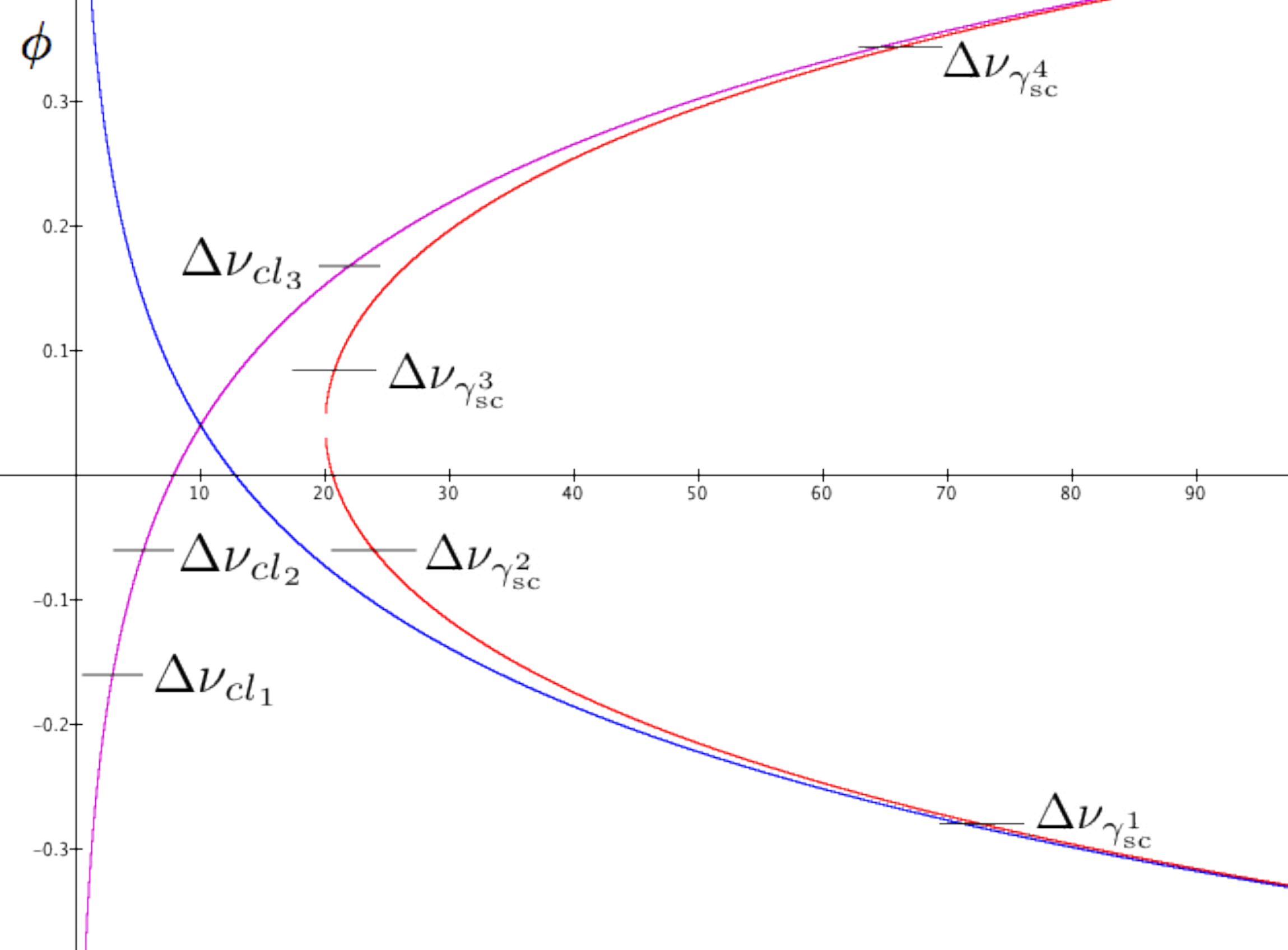}
\label{fig:trajcg}
}%
\subfloat[Coarse-graining by singularity]{
\includegraphics[width=0.485\textwidth]{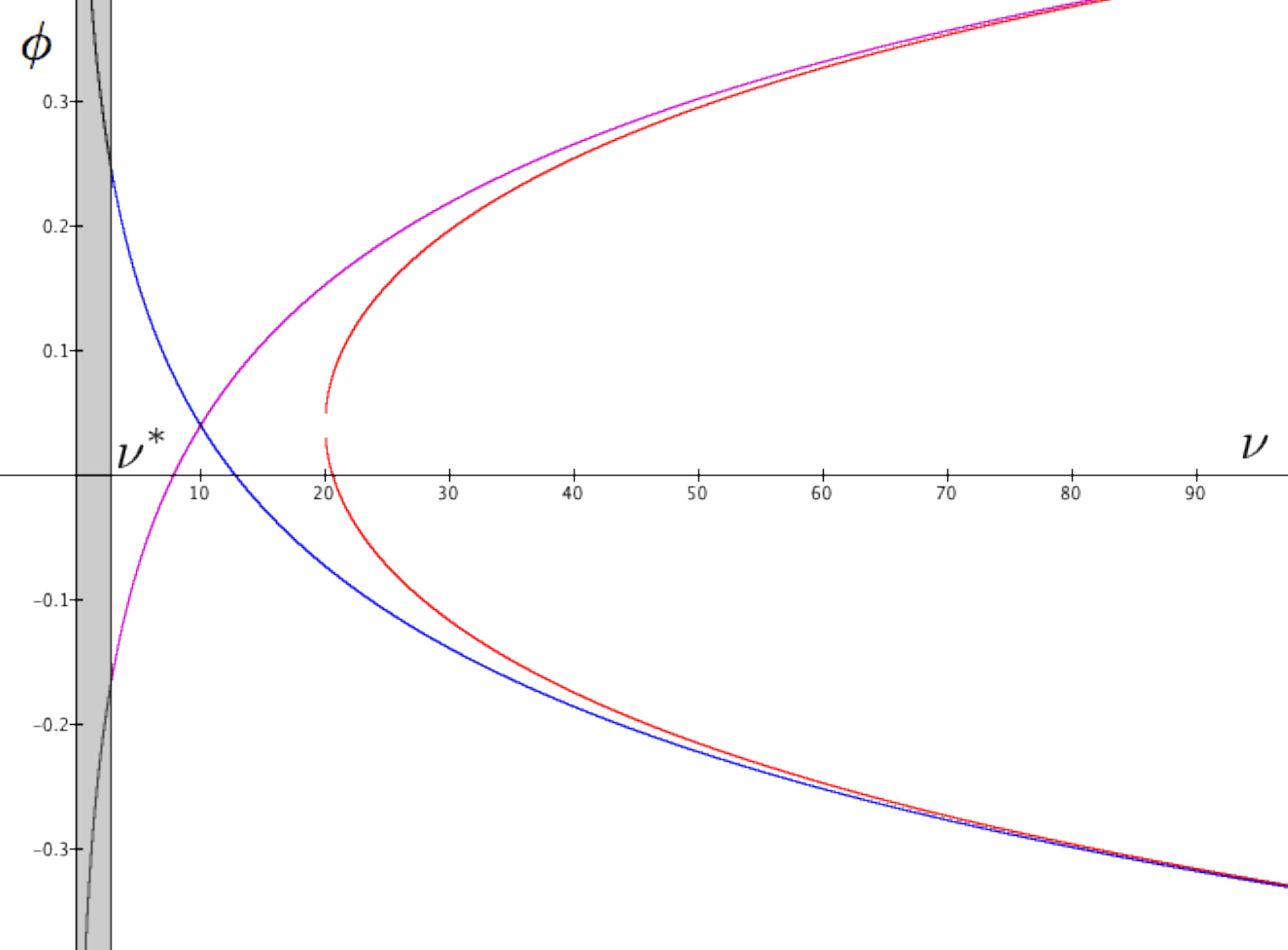}
\label{fig:smallvol}%
}%
\end{center}
\caption{Coarse-grainings of the FLRW minisuperspace.  The left-hand plot
shows an expanding (R-moving) and collapsing (L-moving) classical trajectory
as well as the corresponding ``classical'' loop quantum trajectory i.e.\
solution to the effective Friedmann equation (\ref{eq:LQCFriedmann}).  Also
depicted are coarse-grainings by ranges of values of the volume at different
values of the scalar field for two histories.  The first is a coarse-grained
history $(\Delta\v_{{cl}_1},\Delta\v_{{cl}_2},\Delta\v_{{cl}_3})$ describing a
quantum universe peaked on an expanding classical trajectory.  The second
history
$(\Delta\v_{\gamma_{\text{sc}}^1},\Delta\v_{\gamma_{\text{sc}}^2},\ldots)$
describes a loop quantum trajectory characterized by a bounce, which is peaked
on symmetrically related expanding and collapsing classical trajectories at
large $|\phi|$.
The right-hand plot shows the same trajectories as well as a coarse-graining
suitable for studying the probability that the universe assumes large or small
volume.  The volume $\v$ is partitioned into the range $\Delta\v^*=[0,\v^*]$
(the shaded region in the figure)
and its complement $\overline{\Delta\v^*}=(0,\infty)$.  The quantum universe
may be said to attain small volume if the probability for the branch wave
function $\ket{\Psi_{\smash{\Delta\v^*}}(\phi)}$ is near unity while that for
$\ket{\Psi_{\smash{\overline{\Delta\v^*}}}(\phi)}$ is near zero for arbitrary
choices of $\nu^*$.  Conversely, the universe may be said to attain
arbitrarily large volume over some range of $\phi$ if the probability for
$\ket{\Psi_{\smash{\overline{\Delta\v^*}}}(\phi)}$ is near unity for arbitrary
choice of $\v^*$ over that range of $\phi$.
}%
\label{fig:msstraj}%
\end{figure}

The class operator specifying the volume of the fiducial cell at a sequence of
values of $\phi$ given in (\ref{eq:nutrajclassopdef}) describes a
(coarse-grained) cosmological trajectory $\v(\phi)$.  (See Fig.\
\ref{fig:trajcg} for some examples.)  Because such class operators are not
simply projections, the corresponding branch wave functions will not in
general decohere, and in these simple models probabilities can \emph{not} in
general be assigned to the family of trajectories they describe, as is typical
in quantum theory.  Nonetheless, there are several physically important
examples for which they do decohere.

The simplest way in which decoherence of cosmological trajectories obtains is
if the quantum state remains peaked along a particular individual trajectory,
here imagined as a classical trajectory for convenience -- where ``classical''
means a bouncing solution of the effective Friedmann equation
(\ref{eq:LQCFriedmann}) in the case of sLQC, which approaches the corresponding
general relativistic solutions at large volume.  (See the remarks in Section
\ref{sec:quasiclassical} immediately below.)  Consider a coarse-graining on a
set of slices $\{\phi_1,\phi_2,\ldots,\phi_n\}$ by a set of intervals in
volume $\{\Delta\v_{i_k},k=1\ldots n\}$ chosen in such a way that on each
slice $\phi_k$ one of the ranges $\Delta\v_{cl_k}$ straddles the trajectory.
If $\Delta\v_{cl_k}$ is wider than the width of the quantum state at $\phi_k$,
then essentially the only non-zero branch wave function is
\begin{equation}
  \ket{\Psi_{cl}} =  \Projsupb{\nu}{\Delta\nu_{{cl}_n}}(\phi_n)  \cdots
     \Projsupb{\nu}{\Delta\nu_{{cl}_2}}(\phi_2)
     \Projsupb{\nu}{\Delta\nu_{{cl}_1}}(\phi_1)\ket{\Psi}. 
\label{eq:classtrajbwf}
\end{equation}
The family of histories (classical,non-classical) therefore decoheres and
quasiclassical behavior for such a state is predicted with probability one.
Attempts to specify the trajectory too finely (relative to the width of the
quantum state) destroys the decoherence.  This is important: quantum theory
\emph{has no predictions} for the trajectory followed \emph{even for
quasiclassical states} if it is too precisely specified.  (This is merely the 
uncertainty principle manifesting itself in the context of quantum cosmology.)

A more detailed discussion of what it means for a state to ``follow a
trajectory'' in generalized quantum theory is given in Ref.~\refcite{CS13a}.
A similar but more sophisticated analysis may be given for WKB states; see
Refs.~\refcite{lesH,CH04} for details.

In this review we will describe several important examples of scenarios
involving quantum histories of volume evolving with the scalar field $\phi$.

\subsection{Quasiclassical trajectories}
\label{sec:quasiclassical}

We intend by ``quasiclassical state'' to mean a quantum state that is peaked
on some classical trajectory (\ref{eq:FLRWclasssoln}) at large volume.  (An
explicit example of such a state is given in Eq.~(3.37) of
Ref.~\refcite{CS10c}.)  

We characterize quasiclassical states in this way for
the following reason.  Recall that in the Wheeler-DeWitt quantization, generic
quantum states (\ref{eq:WdWQWFdef}) are a superposition of orthogonal and
independent right- and left-moving (expanding and contracting) branches.  In
loop quantum cosmology, on the other hand, \emph{all} quantum states approach
a specific symmetric superposition of right- and left-moving Wheeler-DeWitt
states at large volume, corresponding to the quantum ``bounce'' of loop
quantum states.

A ``quasiclassical'' state in the Wheeler-DeWitt quantization is one which is 
peaked on either an expanding or contracting classical solution at large 
volume for all values of $\phi$.  In the generic case that the state is a 
superposition of right- and left-moving branches, such a state will be peaked 
on a contracting classical solution as $\phi\rightarrow-\infty$, and an 
expanding solution as $\phi\rightarrow\infty$.  If the state is purely right- 
or left-moving, then only one of these cases will hold.

By contrast, since loop quantum states \emph{always} approach a superposition
of right- and left-moving branches, ``quasiclassical'' states in LQC will be
peaked on a solution of the ``effective'' Friedmann equation
(\ref{eq:LQCFriedmann}), which approaches a classical contracting solution at
large volume as $\phi\rightarrow\-\infty$ and an expanding solution as
$\phi\rightarrow+\infty$, as in the generic Wheeler-DeWitt case.  By the
analysis of states which remain peaked on a trajectory given above, then, a
sufficiently coarse-grained family of trajectory histories will decohere for
such ``quasiclassical'' states, predicting quasiclassical behavior at large
volume with unit probability, so long as the trajectory is not too finely
specified.  (See Refs.~\refcite{CS10c,CS13a} for a more in-depth discussion.)

(Not that this usage therefore does \emph{not} imply that a ``quasiclassical''
state necessarily follows a classical (i.e.\ general relativistic) trajectory
throughout its evolution.  In the Wheeler-DeWitt quantum theory, such states
will in fact track classical solutions all the way to the singularity, while
in the case of sLQC, states which track classical general relativistic
trajectories at large volume are connected by the ``bounce'' in the deep
quantum regime along a solution to the effective Friedmann equation.)

\subsection{Volume singularity}
\label{sec:volsing}

In order to assess whether or not a given quantum cosmological model is 
singular a specific criterion for singularity must be given.  In these simple 
homogeneous isotropic models there are relatively few available, and they are 
clearly connected: does the volume $\v$ of the fiducial cell become 0?  Does 
the curvature $b$ diverge?  Does the matter density $\rho$
(given classically by (\ref{eq:rhodef})) %
diverge?  The latter definition of ``quantum singularity'' seems the most
physical, but it does require adopting a specific definition for the matter
energy density operator.  Several choices are considered in
Ref.~\refcite{acs:slqc}.  For each of these choices essentially the same
underlying physics comes into play: as $p_{\phi}$ is a constant of the motion
in both the classical and the quantum theory, and the density $\rho|_{\phi}$
is essentially a ratio of (the square of) $p_{\phi}$ to (the square of)
volume, $\rho|_{\phi}$ will remain bounded above if the volume remains bounded
below, and will diverge if the volume becomes zero.  These arguments are made
precise in Refs.~\refcite{aps,aps:improved,acs:slqc}, where it is shown that
the expectation value of the volume inevitably becomes 0 for generic states in
the Wheeler-DeWitt theory, and therefore the expectation value of the matter
energy density diverges.  By contrast, in sLQC the expectation value of the
density is bounded above by the critical value (\ref{eq:rhocritdef}) given by
the ``moral'' argument of Ref.~\refcite{dac13a} and Eq.~(\ref{eq:rholim}) due
to the linear scaling of the ultraviolet cutoff on the constraint
eigenfunctions; see Refs.~\refcite{aps,aps:improved,acs:slqc,ashsingh11}, and
Ref.~\refcite{dac13a} for a different proof.

In Refs.~\refcite{CS10c,CS13a} the question of the singularity of the universe
is addressed in the consistent histories framework, in which quantum questions
concerning multiple values of $\phi$ -- i.e.\ concerning the singularity of
quantum \emph{histories} of the universe -- may be framed and confidently
answered.  The most direct course would be to consider ranges of eigenvalues
of the density operator and show that histories for which the density diverges
have probability zero (or unity, in the Wheeler-DeWitt theory.)  However, this
course is not available as the spectrum of the density operator is not
currently known in either theory.

Fortunately, as discussed above, at least in these models the behavior of the 
matter density is directly tied to the behavior of the volume of the fiducial 
cell, and so the analysis of the singularity of these quantum universes can 
instead be given in terms of the volume.%
\footnote{Alternately, an analysis analagous to what we give here can be 
perfomed for the curvature $b$ conjugate to the volume.
} %
We will show that in the Wheeler-DeWitt quantum theory \emph{all} states are
inevitably singular, while in the loop quantization, generic quantum states --
quasiclassical or not -- ``bounce'' from large volume to large volume.  This,
of course, was already known.  However, the consistent histories analysis
brings new, potentially sharper tools to the problem, and sheds additional 
light on the theory, as will be discussed in the conclusion.

The analysis proceeds as follows.  The question of whether the universe
becomes singular is, in these models, equivalent to the question of whether
the volume becomes 0.  Therefore, partition $|\v|$ (recall all quantum states
are symmetric in $\v$) into complementary, disjoint ranges
$\{\Delta\v^*,\overline{\Delta\v^*}\}$, where $\Delta\v^*=[0,\v^*]$ and its
complement $\overline{\Delta\v^*}=(\v^*,\infty)$, where $\v^*$ is any
arbitrary volume.  (See Fig.\ \ref{fig:smallvol}.)  The universe has small
volume at some scalar field value $\phi^*$ if $|\v|\in\Delta\v^*$ at $\phi^*$,
and large volume if $|\v|\in\overline{\Delta\v^*}$.  The corresponding branch
wave functions are
\begin{equation}
\ket{\Psi_{\Delta\v^*|_{\phi^*}}(\phi)}= 
U(\phi-\phi^*)P^{\v}_{\Delta\v^*}\ket{\Psi(\phi^*)}
\label{eq:volbwfdef}
\end{equation}
and its complement; the probability the volume is in $\Delta\v^*$ given the 
state $\ket{\Psi}$ is then
\begin{eqnarray}
p_{\Delta\v^*}(\phi) & =  & 
  \bracket{\Psi_{\Delta\v^*|_{\phi^*}}(\phi)}{\Psi_{\Delta\v^*|_{\phi^*}}(\phi)}
\nonumber\\
 & = & 
\begin{cases}
\int_{0}^{\v^*}\!d\v\, |\psi^{\sWdW}(\v,\phi)|^2 & \text{Wheeler-DeWitt} \\
\sum_{|\v|\in\Delta\v^*} |\psi(\v,\phi)|^2 & \text{sLQC} \ ,
\end{cases}
\label{eq:volprobdef}
\end{eqnarray}
with $p_{\Delta\v^*}(\phi)=1-p_{\overline{\Delta\v^*}}(\phi)$.  These
probabilities are evaluated explicitly for quasiclassical Wheeler-DeWitt and
loop quantum states in Refs.~\refcite{CS10c,CS13a}.  We are interested here,
however, in the question of the singularity of such states.

We begin with the Wheeler-DeWitt theory.  It is shown in detail in
Ref.~\refcite{CS10c} that for purely left- or right-moving states 
\begin{align}
\lim_{\phi\rightarrow-\infty} p^L_{\Delta\nu^*}(\phi) &= 0
&
\lim_{\phi\rightarrow+\infty} p^L_{\Delta\nu^*}(\phi) &= 1
\nonumber\\
\lim_{\phi\rightarrow-\infty} p^R_{\Delta\nu^*}(\phi) &= 1
&
\lim_{\phi\rightarrow+\infty} p^R_{\Delta\nu^*}(\phi) &= 0
\label{eq:pLR}
\end{align}
for any fixed $\v^*$.  This result may be understood most easily by examining 
Eqs.~(\ref{eq:WdWPsiLRdef}).  From the Riemann-Lebesgue lemma it is clear that 
\emph{all} right-/left-moving states are ``sucked in'' to the singularity 
along the classical trajectories (\ref{eq:FLRWclasssoln}).  This means, 
perhaps unsurprisingly, that generic expanding states are inevitably singular 
as $\phi\rightarrow -\infty$, and contracting states are inevitably singular 
as $\phi\rightarrow+\infty$.  Two points merit emphasis.   First, while the 
result is expected, it has been here \emph{derived} within a well-defined 
framework for quantum prediction.  Second, there is the question of the role 
of the limits $|\phi|\rightarrow\infty$.  It may indeed be the case that some 
quantum states become singular at finite $\phi$.  The limit 
$|\phi|\rightarrow\infty$ serves to show that the singularity is a generic 
prediction for \emph{all} states in the theory.

It is noteworthy, however, that generic states in the Wheeler-DeWitt theory 
are actually \emph{superpositions} of expanding and contracting states.  What 
are the corresponding probabilities then?  Indeed, if one writes
\begin{equation}
\ket{\Psi} = \sqrt{p_L}\,\ket{\Psi_L} + \sqrt{p_R}\,\ket{\Psi_R}\ ,
\label{eq:psiLRcatdef}
\end{equation}
one finds
\begin{equation}
p_{\Delta\nu^*}(\phi) = 
  p^L_{\Delta\nu^*}(\phi) + p^R_{\Delta\nu^*}(\phi),
\label{eq:pvolcat}
\end{equation}
with $p^{L,R}_{\Delta\v}(\phi)$ given as above, and
\begin{equation}
\lim_{\phi\rightarrow -\infty} p_{\Delta\nu^*}(\phi) = p_R
\qquad \mathrm{and}\qquad
\lim_{\phi\rightarrow +\infty} p_{\Delta\nu^*}(\phi) = p_L.
\label{eq:pvolcatlim}
\end{equation}
See Fig.\ \ref{fig:pvolplot} for an example (and Ref.~\refcite{CS10c} for
an explicit formula for quasiclassical states.)

By contrast, in loop quantum cosmology one finds\cite{CS13a} instead that, for 
generic states
\begin{align}
\lim_{\phi\rightarrow-\infty} p_{\Delta\nu^*}(\phi) &= 0
&
\lim_{\phi\rightarrow+\infty} p_{\Delta\nu^*}(\phi) &= 0
\nonumber\\
\lim_{\phi\rightarrow-\infty} p_{\overline{\Delta\nu^*}}(\phi) &= 1
&
\lim_{\phi\rightarrow+\infty} p_{\overline{\Delta\nu^*}}(\phi) &= 1.
\label{eq:probvolslim}
\end{align}
from which it is clear that loop quantum states invariably bounce.  The
argument is once again based on the behavior of the eigenfunctions $\eS_k(\v)$
inserted into (\ref{eq:LQCQWFdef}).  The UV cutoff in the eigenfunctions
ensures that for \textbf{any} finite volume $\v^*$, \emph{all} states in sLQC
vanish as $|\phi|\rightarrow\infty$, again on account of Riemann-Lebesgue.  In
addition, essentially because the eigenfunctions vanish at $\v=0$, the
probability the volume of the fiducial cell is precisely 0 is 0 for \emph{all}
$\phi$: $p_{\v=0}(\phi) = 0$.
These universes never become singular for any value of $\phi$.

\subsection{Quantum bounce}
\label{sec:qbounce}

We have shown how, within a consistent histories framework, it may be 
demonstrated that the probability that the universe achieves zero volume is 
zero, while the probability that generic quantum states achieve arbitrarily 
large volume in both limits $|\phi|\rightarrow\infty$ is unity: \emph{all} 
quantum states in sLQC ``bounce'' from large volume to large volume.

By contrast, in the Wheeler DeWitt quantum theory, we showed that all
right-moving (expanding) states inevitably assume zero volume as
$\phi\rightarrow -\infty$, and contracting states do so as $\phi\rightarrow
+\infty$, and are therefore singular.  However, a \emph{superposition} of
expanding and contracting states leads to a probability for the universe
having small volume that is in general never unity, just as it is never 
zero.  This is suggestive that, in the Wheeler-DeWitt quantum theory, such a 
superposition state leads to a non-zero probability that it, too, might 
``bounce'' from large volume to large volume (with probability 
$p_{\mathrm{bounce}}=p_L\cdot p_R$).

This possibility is not realized, however -- \emph{all} quantum states in the
Wheeler-DeWitt quantum theory are singular, just as all states in sLQC are
non-singular.  To understand why this is so it is important to recognize that
the question of whether or not a quantum universe bounces \emph{is not a
question about a single value of $\phi$} -- it is a question about a
coarse-grained \emph{\textbf{trajectory}}: does the universe have a large
volume at (at least) \emph{two} values of $\phi$, one in the ``past'' and one
in the ``future''?  This is a \emph{genuinely quantum} question, and only has
a definite quantum answer in the instance that the corresponding family of
histories decoheres.\cite{CS10a,CS10c,CS13a} We shall show, in fact, that in
an appropriate limit it does, and that indeed loop quantum universes are
generically non-singular and Wheeler-DeWitt universes are singular.

To pose the question concretely, consider partitions of the volume 
$\{\Delta\v^*_1,\overline{\Delta\v^*_1}\}$ and 
$\{\Delta\v^*_2,\overline{\Delta\v^*_2}\}$ on two minisuperspace 
$\phi$-slices $\phi_1$ and $\phi_2$.   The class operator describing a 
``bounce'' is then
\begin{equation}
C_{\mathrm{bounce}}(\phi_1,\phi_2) \,=\,
C_{\overline{\Delta\nu^*_1};\overline{\Delta\nu^*_2}} \, = \, \Projsupb{\nu}{\overline{\Delta\nu^*_1}}(\phi_1)
\Projsupb{\nu}{\overline{\Delta\nu^*_2}}(\phi_2).
\label{eq:Cbouncedef}\\
\end{equation}
The class operator for the complementary history in which the universe is at 
arbitrarily small volume at $\phi_1$, $\phi_2$, or both, is then
\begin{eqnarray}
C_{\mathrm{sing}}(\phi_1,\phi_2)  & = & \Id - C_{\mathrm{bounce}}(\phi_1,\phi_2)
\nonumber\\
& = &
C_{\smash{\Delta\nu^*_1;\Delta\nu^*_2}} +
C_{\smash{\Delta\nu^*_1;\overline{\Delta\nu^*_2}}} +
C_{\smash{\overline{\Delta\nu^*_1};\Delta\nu^*_2} }.
\label{eq:Csingdef}
\end{eqnarray}
By arguments essentially similar to those leading to (\ref{eq:pLR}) 
above, one shows that in the Wheeler-DeWitt quantum theory the branch wave 
functions corresponding to bouncing vs.\ singular cosmologies are
\begin{eqnarray} 
\ket{\Psi_{\text{sing}}(\phi)}  & = & 
U(\phi-\phi_o) \lim_{\substack{\phi_1\rightarrow -\infty\\ \phi_2\rightarrow +\infty }}
C_{\mathrm{sing}}^\dagger(\phi_1,\phi_2)\ket{\Psi} = \ket{\Psi(\phi)}
\nonumber\\
\ket{\Psi_{\text{bounce}}(\phi)}  & = & 
 U(\phi-\phi_o)  \lim_{\substack{\phi_1\rightarrow -\infty\\ 
\phi_2\rightarrow +\infty }} C_{\text{bounce}}^{\dagger} (\phi_1,\phi_2) \ket{\Psi} 
  = 0 ~.
\label{eq:bwfdefb-s}
\end{eqnarray}
Thus the alternative histories (bounce,singular) \emph{decohere} in this
limit, $d(\mathrm{bounce},\mathrm{singular})=
\bracket{\Psi_{\mathrm{sing}}}{\Psi_{\mathrm{bounce}}}=0$, and
\begin{eqnarray}
p_{\mathrm{sing}} & = & \bracket{\Psi_{\mathrm{sing}}}{\Psi_{\mathrm{sing}}}
\nonumber\\
 & = & \bracket{\Psi}{\Psi}
\nonumber\\
 & = & 1.
\label{eq:psing}
\end{eqnarray}
Wheeler-DeWitt quantum cosmological models are \emph{invariably} singular,
regardless of state, in spite of the potential promise of Fig.\
\ref{fig:pvolplot} that they sometimes might not be.

Similarly, in loop quantum cosmology one finds instead that 
\begin{eqnarray}
\ket{\Psi_{\mathrm{bounce}}} & = & C_{\mathrm{bounce}}\ket{\Psi}
\nonumber\\
 & = & \ket{\Psi}
\label{eq:psibounce}
\end{eqnarray}
while
\begin{eqnarray}
\ket{\Psi_{\mathrm{sing}}} & = & C_{\mathrm{sing}}\ket{\Psi}
\nonumber\\
 & = & 0.
\label{eq:psising}
\end{eqnarray}
Again, these histories decohere, but now $p_{\mathrm{bounce}} =1$,
and the probability arbitrary loop quantum states are singular is 0.

\section{Discussion}
\label{sec:disc}

We have here reviewed the formulation of a consistent histories approach to
quantum theory for loop quantum cosmology, showing how the framework provides
the structure necessary -- the decoherence functional -- to enable the theory
to make consistent quantum predictions in the absence of measurements,
external observers, or other similar apparatus typically required in quantum
theory before one can make definite predictions.  We have illlustrated the
application of the framework in showing how loop quantum cosmologies are
non-singular for generic states, contrasting that striking result with the
inevitable singularity of the Wheeler-DeWitt quantization of the same family
of comological models.  We have also made an effort to point to some other
work on consistent histories formulations of quantum cosmological models, even
though there was not sufficient space to review them here.

Discussion of a few important points is in order.  It is no secret that while
physicists generally agree on how to \emph{do} quantum mechanics, the story is
quite different when one has the temerity to inquire what it \emph{means}.  We
wish to emphasize that it is not necessary to commit to any particular
ontology for quantum mechanics to acknowledge the centrality of the role of
interference among alternative outcomes in arriving at quantum predictions.
Indeed, destruction of this interference is the fundamental \emph{technical}
role played by the classical idea of ``measurement'' in quantum theory via the
postulated ``collapse'' of the wave function upon measurement of a property by
an external agent.  Consistent histories quantum theory supplies, more or
less, the \textbf{minimal} additional structure one requires to provide
quantum theory with an objective, observer independent, purely \emph{internal}
measure of this interference, that \emph{reproduces the predictions of
ordinary measurement-based quantum theory} in traditional ``measurement
situations'', but also \emph{extends it} to physical circumstances in which
there is no meaningful notion of an ``external observer'', thus enabling
quantum theory to make predictions concerning questions of fundamental
physical interest such as in the early universe.  In particular, we emphasize
that apart from introducing the decoherence functional, the objective measure
of interference derived from the quantum state (an object already present in
the conventional theory), consistent histories quantum theory \emph{is still
quantum theory}, with all of the interpretational challenges that implies.%
\footnote{Some of the foundational questions the consistent histories
framework does not by itself attempt to directly answer include the problem of
outcomes, the basis problem, and the meaning of ``probability'' -- among
others.
} %
It is neither the place nor our intention here to analyze the limitations of
the consistent histories framework as a complete answer to the question,
``What is the quantum mechanical account of reality?''  It is our view that
while the concepts of consistent histories -- even supplemented by physical
mechanisms and ideas such as environmental decoherence, envariance, and
``quantum Darwinism''\cite{giulini,schlosshauer07,zurek09a,RZZ16a} -- are at
best only a \emph{partial} answer to this question, we nonetheless believe it
likely they will play a role in an eventual picture.

We have just alluded to the fact that the consistent histories framework does
not offer a fresh answer to the ``true'' meaning of the probabilities quantum
theory supplies.  Nonetheless, we take it as uncontroversial that the meaning
of a probability that is unity or zero is not ambiguous -- the theory predicts
that thing either does, or does not, happen, with
certainty.\cite{hartle88a,hartle91a,sorkin94,sorkin97a} These are the
\textbf{definite} predictions of a theory.  It is for this reason we have
emphasized the $|\phi|\rightarrow\infty$ limit in our assessment of the
probabilities that Wheeler-DeWitt or loop quantum universes are (or are not)
singular.  It is in this limit that we are \emph{guaranteed} that \textbf{all}
Wheeler-DeWitt states are singular, and that \textbf{all} loop quantum states
bounce, quasiclassical or not.

It may be natural to inquire whether one could have argued that Wheeler-DeWitt
cosmologies are necessarily singular even for superposition states such as
(\ref{eq:psiLRcatdef}) by arguing that the amplitude
$\bracket{\Psi_L}{\Psi_R}=0$, without invoking the consistent histories
formulation, and in spite of the fact that the ``single-$\phi$'' probability
illustrated in Fig.\ \ref{fig:pvolplot} suggests otherwise.  (Indeed, one
finds such arguments in the classical literature on quantum
cosmology.\cite{halliwell91:qcbu,kiefer12}) This is tempting, and is certainly
the thrust of the consistent histories calculation itself.  However, doing so
\emph{assumes} that this amplitude may be interpreted as a probability for a
sequence of events, and thereby glosses over one of the fundamental messages
of quantum theory: \emph{\textbf{quantum amplitudes do not represent quantum
probabilities unless interference among the alternative outcomes vanishes.}}%
\footnote{It may be worth reiterating that this obtains generally when one
considers amplitudes for \emph{sequences} of quantum events -- such as which
slit a particle passes through in a two slit apparatus.  But such amplitudes
for quantum histories are precisely the sort in which one is often interested
in quantum cosmology, such as the amplitudes for a quantum bounce discussed in
this review.\cite{CS10a,CS10b,CS10c,CS13a}
} %
Simply put, quantum mechanics says that \textbf{some questions do not have
physically meaningful answers.} In ordinary laboratory applications of quantum
mechanics, the act of gathering information about a system -- ``measurement''
-- supplies the physical mechanisms that destroy that interference.
Amplitudes for outcomes which are measured do not interfere, and therefore may
be interpreted as probabilities for those measured outcomes.  If the
measurement is not made, those amplitudes may \textbf{not} be consistently
interpreted as probabilities for the unmeasured outcomes.  The consistent
histories perspective on quantum theory simply recognizes that amplitudes are
not probabilities unless such interference vanishes, and supplies an objective
measure of that interference that may be applied in the absence of
laboratories and measurements in environments such as the very early universe,
where there were certainly no observers present.  Even so, physical mechanisms
may exist which destroy interfence among possible alternatives, and thereby
enable quantum theory to assign definite probabilities to those alternatives.
Indeed, additional degrees of freedom may supply a resource which effectively
``gathers information'' (i.e.\ creates
records\cite{GMH90a,GMH90b,hartle91a,GMH93,halliwell99,RZZ16a}) about the
physical alternatives of interest, leading to decoherence of those alternative
histories and consequent ability to assign them meaningful probabilities.  One
would expect this to be the case in the actual physical universe which even in
a globally homogeneous and isotropic cosmology carries both geometric and
matter degrees of freedom which may imprint information about quantum
alternatives and lead to decoherence.  Matter density perturbations in the
early universe are just such degrees of freedom, and a consistent histories
analysis of such perturbations should lead to a coherent picture of the
``quantum-to-classical'' transition of inflationary perturbations.\cite{CS17a}

Thus, the methods described in this review provide the tools to tell a
consistent quantum story of the predictions of quantum gravitational theories
in the early universe.  In forthcoming work these methods will be applied to
provide a similar analysis of spin-foam loop quantum cosmological
models.\cite{CS13b}


\section*{Acknowledgments}

The author would like to thank P.\ Singh for teaching him loop quantum 
cosmology and for a fruitful collaboration.  Portions of this work were 
supported by a grant from FQXi, for which we thank the Institute.


%


\ifthenelse{\arxiv=1}{%
\bibliography{dXXXrev}
}{%
\bibliographystyle{ws-ijmpd}
\bibliography{global_macros,../Bibliographies/master}%

\begin{thebibliography}{10}

\bibitem{griffiths08}
R.~B. Griffiths, {\em Consistent quantum theory} (Cambridge University Press,
  Cambridge, 2008).

\bibitem{omnes94}
R.~Omn\`es, {\em The interpretation of quantum mechanics} (Princeton University
  Press, Princeton, 1994).

\bibitem{GMH90a}
M.~Gell-Mann and J.~B. Hartle, Quantum mechanics in the light of quantum
  cosmology, in {\em Proceedings of the 3rd international symposium on the
  foundations of quantum mechanics in the light of new technology\/},  eds.
  S.~Kobayashi, H.~Ezawa, M.~Murayama and S.~Nomura (Physical Society of Japan,
  Tokyo, 1990) pp. 321--343.

\bibitem{GMH90b}
M.~Gell-Mann and J.~B. Hartle, Quantum mechanics in the light of quantum
  cosmology, in {\em Complexity, Entropy, and the Physics of Information\/},
  ed. W.~Zurek, SFI Studies in the Sciences of Complexity, Vol.~VII
  (Addison-Wesley, Reading, 1990) pp. 425--458.

\bibitem{hartle91a}
J.~B. Hartle, The quantum mechanics of cosmology  in Coleman {\em
  et~al.}\cite{qcbu} pp. 65--157.

\bibitem{lesH}
J.~B. Hartle, Spacetime quantum mechanics and the quantum mechanics of
  spacetime, in {\em Gravitation and Quantizations, Proceedings of the 1992 Les
  Houches Summer School\/},  eds. B.~Julia and J.~Zinn-Justin (North Holland,
  Amsterdam, 1995) pp. 285--480.

\bibitem{halliwell99}
J.~J. Halliwell, {\em Phys.\ Rev.} {\bf D60}  (1999) 105031,
  \href{http://arxiv.org/abs/quant-ph/9902008}{{\ttfamily
  arXiv:quant-ph/9902008 [quant-ph]}}.

\bibitem{hallithor01}
J.~J. Halliwell and J.~Thorwart, {\em Phys.\ Rev.} {\bf D64}  (2001)   124018,
  \href{http://arxiv.org/abs/gr-qc/0106095}{{\ttfamily arXiv:gr-qc/0106095
  [gr-qc]}}.

\bibitem{hallithor02}
J.~J. Halliwell and J.~Thorwart, {\em Phys.\ Rev.} {\bf D65}  (2002)   104009,
  \href{http://arxiv.org/abs/gr-qc/0201070}{{\ttfamily arXiv:gr-qc/0201070
  [gr-qc]}}.

\bibitem{halliwall06}
J.~J. Halliwell and P.~Wallden, {\em Phys.\ Rev.} {\bf D73}  (2006)   024011,
  \href{http://arxiv.org/abs/gr-qc/0509013}{{\ttfamily arXiv:gr-qc/0509013
  [gr-qc]}}.

\bibitem{halliwell09}
J.~J. Halliwell, {\em Phys.\ Rev.} {\bf D80}  (2009)   124032,
  \href{http://arxiv.org/abs/0909.2597}{{\ttfamily arXiv:0909.2597 [gr-qc]}}.

\bibitem{hartlemarolf97}
J.~B. Hartle and D.~Marolf, {\em Phys.\ Rev.} {\bf D56}  (1997) 6247,
  \href{http://arxiv.org/abs/gr-qc/9703021}{{\ttfamily arXiv:gr-qc/9703021
  [gr-qc]}}.

\bibitem{CH04}
D.~A. Craig and J.~B. Hartle, {\em Phys. Rev.} {\bf D69} (June 2004) 123525,
  \href{http://arxiv.org/abs/gr-qc/0309117v3}{{\ttfamily arXiv:gr-qc/0309117v3
  [gr-qc]}}.

\bibitem{as05}
C.~Anastopoulos and K.~Savvidou, {\em Class.\ Quant. Grav.} {\bf 22}  (2005)
  1841, \href{http://arxiv.org/abs/gr-qc/0410131}{{\ttfamily
  arXiv:gr-qc/0410131 [gr-qc]}}.

\bibitem{giulini}
E.~Joos, H.~D. Zeh, C.~Kiefer, D.~Giulini, J.~Kupsch and I.-O. Stamatescu, {\em
  Decoherence and the appearance of a classical world in quantum theory},
  second edn. (Springer-Verlag, Berlin, 2003).

\bibitem{schlosshauer07}
M.~Schlosshauer, {\em Decoherence and the quantum-to-classical transition}
  (Springer-Verlag, Berlin, 2007).

\bibitem{halliwell89}
J.~J. Halliwell, {\em Phys. Rev.} {\bf D39}  (1989)   2912.

\bibitem{zurek09a}
W.~H. Zurek, {\em Nature Physics} {\bf 5}  (2009) 181,
  \href{http://arxiv.org/abs/0903.5082}{{\ttfamily arXiv:0903.5082
  [quant-ph]}}.

\bibitem{RZZ16a}
C.~J. Riedel, W.~H. Zurek and M.~Zwolak, {\em Phys. Rev.} {\bf A93}  (2016)
  032126, \href{http://arxiv.org/abs/1312.0331}{{\ttfamily arXiv:1312.0331
  [quant-ph]}}.

\bibitem{CS10a}
D.~A. Craig and P.~Singh, {\em Found.\ Phys.} {\bf 41}  (2011) 371,
  \href{http://arxiv.org/abs/1001.4311}{{\ttfamily arXiv:1001.4311 [gr-qc]}}.

\bibitem{CS10b}
D.~A. Craig and P.~Singh, A consistent histories formulation of
  {Wheeler-DeWitt} quantum cosmology, in {\em Quantum Theory: Reconsideration
  of Foundations -- 5\/},  ed. A.~Krennikhov (American Institute of Physics,
  New York, 2010) pp. 275--282.
\newblock Proceedings of the fifth {V\"{a}xj\"{o}} conference on the
  foundations of quantum mechanics, 14-18 June 2009.

\bibitem{CS10c}
D.~A. Craig and P.~Singh, {\em Phys. Rev.} {\bf D82}  (2010) 123526,
  \href{http://arxiv.org/abs/1006.3837}{{\ttfamily arXiv:1006.3837 [gr-qc]}}.

\bibitem{CS12a}
D.~A. Craig and P.~Singh, {\em J. Phys.: Conf. Series} {\bf 360}  (2012)
  012028.

\bibitem{CS13a}
D.~A. Craig and P.~Singh, {\em Class. Quantum Grav.} {\bf 30}  (2013)   205008,
  \href{http://arxiv.org/abs/1306.6142}{{\ttfamily arXiv:1306.6142 [gr-qc]}}.

\bibitem{CS16c}
D.~A. Craig and P.~Singh, Consistent probabilities in spin foam loop quantum
  cosmology  (2016), in preparation.

\bibitem{GMH93}
M.~Gell-Mann and J.~B. Hartle, {\em Phys.\ Rev.} {\bf D47}  (1993) 3345,
  \href{http://arxiv.org/abs/gr-qc/9210010}{{\ttfamily arXiv:gr-qc/9210010
  [gr-qc]}}.

\bibitem{diosi04}
L.~Diosi, {\em Phys.\ Rev.\ Lett.} {\bf 92}  (2004)   170401,
  \href{http://arxiv.org/abs/quant-ph/0310181}{{\ttfamily
  arXiv:quant-ph/0310181 [quant-ph]}}.

\bibitem{ILS94a}
C.~J. Isham, N.~Linden and S.~Schreckenberg, {\em J. Math. Phys.} {\bf 35}
  (1994) 6360.

\bibitem{dac97}
D.~A. Craig, The geometry of consistency: decohering histories in generalized
  quantum theory  (1997),
  \href{http://arxiv.org/abs/quant-ph/9704031}{{\ttfamily
  arXiv:quant-ph/9704031 [quant-ph]}}.

\bibitem{CS13b}
D.~A. Craig and P.~Singh, Consistent probabilities in spin foam loop quantum
  cosmology  (2016), in preparation.

\bibitem{schroer13a}
D.~P. Schroeren, {\em Found. Phys.} {\bf 43}  (2013) 310,
  \href{http://arxiv.org/abs/1206.4553}{{\ttfamily arXiv:1206.4553 [gr-qc]}}.

\bibitem{ashsingh11}
A.~Ashtekar and P.~Singh, {\em Class. Quantum Grav.} {\bf 28}  (2011)   213001,
  \href{http://arxiv.org/abs/1108.0893}{{\ttfamily arXiv:1108.0893 [gr-qc]}}.

\bibitem{dac13a}
D.~A. Craig, {\em Class. Quantum Grav.} {\bf 30}  (2013)   035010,
  \href{http://arxiv.org/abs/1207.5601}{{\ttfamily arXiv:1207.5601 [gr-qc]}}.

\bibitem{acs:slqc}
A.~Ashtekar, A.~Corichi and P.~Singh, {\em Phys. Rev.} {\bf D77}  (2008)
  024046, \href{http://arxiv.org/abs/0710.3565}{{\ttfamily arXiv:0710.3565
  [gr-qc]}}.

\bibitem{aps}
A.~Ashtekar, T.~Pawlowski and P.~Singh, {\em Phys. Rev.} {\bf D73}  (2006)
  124038, \href{http://arxiv.org/abs/gr-qc/0604013}{{\ttfamily
  arXiv:gr-qc/0604013 [gr-qc]}}.

\bibitem{aps:improved}
A.~Ashtekar, T.~Pawlowski and P.~Singh, {\em Phys. Rev.} {\bf D74}  (2006)
  084003, \href{http://arxiv.org/abs/gr-qc/0607039}{{\ttfamily
  arXiv:gr-qc/0607039 [gr-qc]}}.

\bibitem{dgs14a}
P.~Diener, B.~Gupt and P.~Singh, {\em Class. Quantum Grav.} {\bf 31}  (2014)
  105015, \href{http://arxiv.org/abs/1402.6613}{{\ttfamily arXiv:1402.6613
  [gr-qc]}}.

\bibitem{dgms14a}
P.~Diener, B.~Gupt, M.~Megevand and P.~Singh, {\em Class. Quantum Grav.} {\bf
  31}  (2014)   165006, \href{http://arxiv.org/abs/1406.1486}{{\ttfamily
  arXiv:1406.1486 [gr-qc]}}.

\bibitem{ach09}
A.~Ashtekar, M.~Campiglia and A.~Henderson, {\em Phys. Lett.} {\bf B681}
  (2009) 347.

\bibitem{ach10a}
A.~Ashtekar, M.~Campiglia and A.~Henderson, {\em Class. Quantum Grav.} {\bf 27}
   (2010)   135020, \href{http://arxiv.org/abs/1001.5147v2}{{\ttfamily
  arXiv:1001.5147v2 [gr-qc]}}.

\bibitem{ach10b}
A.~Ashtekar, M.~Campiglia and A.~Henderson, {\em Phys. Rev.} {\bf D82}  (2010)
   124043, \href{http://arxiv.org/abs/1011.1024}{{\ttfamily arXiv:1011.1024
  [gr-qc]}}.

\bibitem{singh06a}
P.~Singh, {\em Phys. Rev.} {\bf D73}  (2006)   063508,
  \href{http://arxiv.org/abs/gr-qc/0603043}{{\ttfamily arXiv:gr-qc/0603043
  [gr-qc]}}.

\bibitem{taveras08a}
V.~Taveras, {\em Phys. Rev.} {\bf D78}  (2008)   064072,
  \href{http://arxiv.org/abs/0807.3325}{{\ttfamily arXiv:0807.3325 [gr-qc]}}.

\bibitem{hrvwe11}
A.~Henderson, C.~Rovelli, F.~Vidotto and E.~Wilson-Ewing, {\em Class. Quantum
  Grav.} {\bf 28}  (2011)   025003,
  \href{http://arxiv.org/abs/1010.0502v2}{{\ttfamily arXiv:1010.0502v2
  [gr-qc]}}.

\bibitem{dac07}
D.~A. Craig, Branch wave functions for quasi-classical homogeneous universes,
  in {\em Proceedings of the Eleventh Marcel Grossmann Meeting on General
  Relativity and Relativistic Astrophysics\/},  ed. R.~Ruffini (World
  Scientific, Singapore, 2007) pp. 1884--1886.

\bibitem{CS16a}
D.~A. Craig and P.~Singh, The vertex expansion in the consistent histories
  formulation of spin foam loop quantum cosmology, in {\em Proceedings of the
  Fourteenth Marcel Grossmann Meeting on General Relativity and Relativistic
  Astrophysics\/},  ed. R.~Ruffini (World Scientific, Singapore, 2016)
\newblock to appear.

\bibitem{CS16b}
D.~A. Craig and P.~Singh, {\em Class. Quantum Grav.}   (2016) to appear.

\bibitem{fpps12}
F.~Falciano, R.~Pereira, N.~Pinto-Neto and E.~S. Santini, {\em Phys. Rev.} {\bf
  D86}  (2012)   063504, \href{http://arxiv.org/abs/1206.4021}{{\ttfamily
  arXiv:1206.4021 [gr-qc]}}.

\bibitem{pf13a}
N.~Pinto-Neto and J.~Fabris, {\em Class. Quantum Grav.} {\bf 30}  (2013)
  143001, \href{http://arxiv.org/abs/1306.0820}{{\ttfamily arXiv:1306.0820
  [gr-qc]}}.

\bibitem{halliwell91:qcbu}
J.~J. Halliwell, Introductory lectures on quantum cosmology  in Coleman {\em
  et~al.}\cite{qcbu} pp. 159--243.

\bibitem{hartle88a}
J.~B. Hartle, Quantum cosmology, in {\em Highlights in gravitation and
  cosmology\/},  eds. B.~Iyer, A.~Kembhavi, J.~V. Narlikar and C.~Vishveshwara
  (Cambridge University Press, Cambridge, 1988)

\bibitem{sorkin94}
R.~Sorkin, {\em Mod. Phys. Lett.} {\bf A9}  (1994) 3119,
  \href{http://arxiv.org/abs/gr-qc/9401003}{{\ttfamily arXiv:gr-qc/9401003
  [gr-qc]}}.

\bibitem{sorkin97a}
R.~Sorkin, Quantum measure theory and its interpretation, in {\em
  Quantum-classical correspondence: Proceedings of the 4th {Drexel Symposium on
  Quantum Nonintegrability}\/},  eds. D.~Feng and B.-L. Hu (International
  Press, Cambridge, Massachusetts, 1997) pp. 229--251.

\bibitem{kiefer12}
C.~Kiefer, {\em Quantum gravity}, third edn. (Oxford University Press, Oxford,
  2012).

\bibitem{CS17a}
D.~A. Craig and P.~Singh, The quantum-to-classical transition of perturbations
  in quantum cosmology  (2017), in preparation.

\bibitem{qcbu}
S.~Coleman, J.~B. Hartle, T.~Piran and S.~Weinberg (eds.), {\em Quantum
  cosmology and baby universes: Proceedings of the 1989 {Jerusalem} Winter
  School for Theoretical Physics} (World Scientific, Singapore, 1991).

\end{thebibliography}
}%

\end{document}